\documentstyle[12pt,epsf]{article}
\begin{document}

\begin{center}

{\Large Collective Transport in Random Media: From
Superconductors to Earthquakes}
\bigskip
\bigskip

Lectures at Summer School on\\
``Fundamental Problems in Statistical Mechanics
IX''\\
August 20-23, 1997\\

\bigskip

Daniel S. Fisher\\
Physics Department, Harvard University\\
Cambridge, MA 02138\\
fisher@cmt.harvard.edu
\end{center}

\bigskip

\begin{abstract}

In these lectures, a variety of non-equilibrium transport phenomena are 
introduced that all involve, in some way, elastic manifolds being driven through 
random media. A simple class of models is studied focussing on the behavior 
near to the  critical ``depinning" force above which  persistent motion
occurs in 
these systems. A simple mean field theory and a ``toy" model of
``avalanche" 
processes are analyzed and used to motivate the general scaling picture found in 
recent renormalization group studies.  The general ideas and results are then 
applied to various systems: sliding charge density waves, critical current 
behavior of vortices in superconductors, dynamics of cracks, and simple models 
of a geological fault. The roles of thermal fluctuations, defects, inertia, and 
elastic wave propagation are all discussed briefly.
\end{abstract}

\subsection*{I. Introduction}

Many phenomena in nature involve transport of
material or some other quantity
from one region of space to another.  In some
cases transport occurs in systems that are close
to equilibrium with the transport representing
only a small perturbation such as flow or
electrical current in a metal, while in other
cases it involves systems that are far from
equilibrium such as  a landslide down a mountain,
or a drop of water sliding down an irregular
surface. Sometimes, particles or other
constituents move relatively independently of
each other like the electrons in a metal, while
in other situations the interactions play an
important role, as in the landslide and the water
droplet.  

If the interactions are strong enough,
all the particles (or other constituents) move
together and the macroscopic dynamics involves
only a small number of degrees of freedom.  This
is the case for a small water drop on, e.g. wax
paper, which slides around while retaining its
shape.  But if the interactions are not so
strong  relative to the other forces acting on
the constituents, then the transport involves in
an essential way {\it many interacting degrees
of freedom}.  This is the case for a  larger
water drop on an irregular surface for which the
contact line between the droplet and the surface
continually deforms and adjusts its shape in
response to the {\it competition} between the
surface tension of the water and the
interactions with the substrate
\cite{ConL}\cite{EK}.  Such a moving drop and  the
landslide  are  examples of non-equilibrium {\it
collective transport phenomena}, which will be
the general subject of these lectures.  

This is, of course, an impossibly
broad subject! We must thus narrow the scope
drastically.  Although the range of systems
discussed here will, nevertheless, be reasonably
broad, we will primarily focus on systems in
which the interactions are strong enough so that
the transported object (or at least some
part of it) is {\it elastic}.  We
will use this in a general and somewhat loose
sense that the transported object has enough
integrity that if one part of it moves a long
distance then so, eventually, must the other
parts as well.  Thus the fluid drop is elastic if
it does not break up---i.e. its perimeter retains
its integrity---while a landslide is not elastic
as some rocks will fall much further than others
and the relative positions of the rocks will be
completely jumbled by the landslide.

We will be interested in systems in which the
medium in which the transport occurs has static
random heterogeneities (``quenched randomness'')
which exert forces on the transported object
that depend on where it is in space.  

Examples we will discuss are: interfaces between
two phases in random
media \cite{Int1}\cite{Int2}, such as between two
fluids in a porous medium \cite{PorE}, or domain
walls in a random ferromagnetic alloy; lattices of
vortices in dirty type II superconductors
\cite{Blat}; charge density waves which are
spatially periodic modulations of the electron
density that occur in certain solids
\cite{CDW1}\cite{CDWF}\cite{CDWNF}; and the motion
of geological faults \cite{EQ1},\cite{EQ2}.  In
addition to the contact line of the fluid drop
already mentioned \cite{ConL},\cite{EK}, another
well known---but poorly understood---example that
we will, however, not discuss is solid-on-solid
friction.

In all of these systems, a driving force, call
it $F$, can be applied which acts to try to make
the object move, but this will be resisted by
the random ``pinning'' forces exerted by the
medium or substrate. The primary questions of
interest will involve the response of the system
to such an applied driving force \cite{FFF}.  If $F$ is
small, then one might guess that it will not be
sufficient to overcome the resistance of the
pinning forces; sections of the object would
 just move a bit and it would deform in response
to $F$, but  would afterwords be at rest.  [Note
that in most of what follows, we will ignore
fluctuations so that the motion is deterministic
and the objects can be said to be stationary.]  If
the force is increased, some segment might go
unstable and move only to be stopped by
stronger pinning regions or neighboring
segments.  But for large enough
$F$, it should be possible to overcome the pinning
forces---unless they are so strong that the
object is broken up, an issue we will return to
at the end---and the object will move, perhaps
attaining some steady state velocity $v$.  

Basic
questions one might ask are: is there a {\it
unique}, history independent force,
$F_c$ separating the static from the moving
regimes?  How does $v$ depend on
$F$ (and possibly on history)? Are there some
kinds of non-equilibrium critical phenomena when
$v$ is small? How does the system respond to an
additional time or space dependent applied force?
These are all {\it macroscopic} properties of
the system.  

But we will also be interested in
some {\it microscopic} properties: how can one
characterize (statistically) the deformations of
the object when it is stationary \cite{BTW}?  The dynamic
deformations and local velocities when it is
moving? The response to a small local
perturbation? etc.

Motivated by possible analogies with equilibrium
phase transitions \cite{HohH}, we can ask if there
are {\it scaling laws} that might obtain near a
critical force which relate, for example, the
characteristic length scale $L$, for some
process, to its characteristic time scale,
$\tau$, via a power law relation of the form:
\begin{equation}
\tau\sim L^z
\label{eq:a1}
\end{equation}

Trying to answer some of these questions---and
to pose other more pointed questions---is the main
aim of these lectures.  In the next few sections 
a particular system and its natural (theoretical)
generalizations will be studied and tools and
ideas developed.  In the last section, these are
tentatively applied to various physical systems
and some of the complicating features left out of
the simple model systems are discussed. This leads
naturally to many open questions.

\subsection*{II. Interfaces and Models}

In order to develop some of the general
ideas---both conceptual and computational ---we
will focus initially on an
{\it interface} between two phases that is driven by
an applied force through an inhomogenous
medium \cite{Int1},\cite{Int2}.  The essential
ingredients of a model of this system are:  the forces
of sections of the interface on nearby sections, i.e.
the elasticity of the interface caused by its
interfacial tension; the preference of the interface
for some regions of the system over others due to
the random heterogeneities; and some dynamical
law which governs the time evolution of the
local interface position.  

\begin{figure}
  \begin{center}
    \leavevmode
    \epsfxsize=3.5truein
    \epsfbox{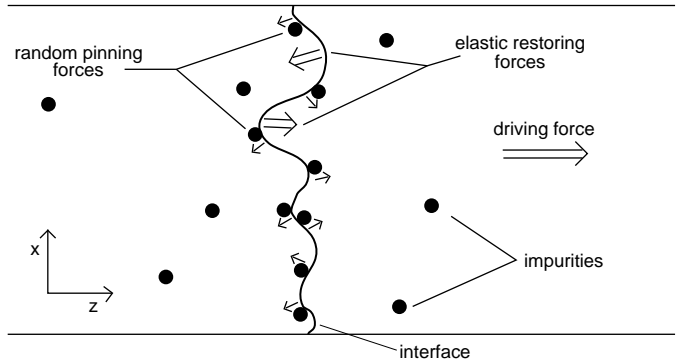}
  \end{center}
\caption{Schematic of a one-dimensional
interface in a two dimensional disordered system
illustrating the forces acting on the interface.}
  \label{fig:1}
\end{figure}

We will initially
make several simplifying approximations, which
we will come back and examine later.  First, we
assume that the interface is not too distorted
away from a flat surface normal to the direction
($z$) of the driving force so that its
configuration can be represented by its
displacement field 
$u(\vec r)$ away from a flat
reference surface. The coordinates $\vec
R=(x,y,z)$ of points on the surface are then
\begin{equation}
(x,y)=\vec r\ \mbox{ and }\  z=u(\vec r).
\label{eq:B1}
\end{equation} 
Second, we will assume that the dynamics are
purely {\it dissipative} i.e. that inertia is
negligible---a good approximation in many physical
situations.  Keeping only the lowest order
terms in deviations from flat, we then have
\begin{equation}
\eta\frac{\partial u(\vec r, t)}{\partial t} =
F+\sigma(\vec r, t)-f_p \left[\vec r,
u(\vec r,t) \right]
\label{eq:B2}
\end{equation} 
with $F$ representing the driving force on the
interface, $f_p(\vec R)$ representing the random
``pinning'' forces of the heterogeneous medium
on the interface which we assume for the present
are not history or velocity dependent; $\eta$ 
a dissipative coefficient; and 
\begin{equation}
\sigma (\vec r, t) = \int d\vec r\,' \int^t dt'
 J (\vec r - \vec r\,', t-t')\left[ u(\vec r\,',
t')-u(\vec r,t) \right]
\label{eq:B3}
\end{equation}
 the ``stress'' on the interface from its
elasticity which is ``transmitted'' by the
kernel $J(\vec r,t)$. Short range elasticity
of the interface corresponds to
\begin{equation}
J\alpha\delta (t) \nabla^2\delta(\vec r\,).
\label{eq:B4}
\end{equation}
A schematic of such an interface and the forces
acting on it is shown in Fig~\ref{fig:1}.

Keeping in mind some of the other problems of
interest \cite{ConL},\cite{Ram1,Ball} in addition to
interfaces, we will abstract to a more general problem
of a
$d$-dimensional elastic ``manifold''---$d=2$ for
the interface---with more general
interactions, which can be 
long-range, embodied in $J(\vec r, t)$.
In addition to the form of $J(\vec r, t)$, the
system will be characterized by the statistics
of the pinning forces which impede interface
motion near points where the interface has
lower (free) energy; $f_p(\vec R)$ will
generally have only short-range correlations in
space, i.e., in both $u-u'$ and $\vec r-\vec
r'$.  Even with these simplifying
assumptions, the model
Eqs~(\ref{eq:B2}, \ref{eq:B3}) is impossible to
analyze fully due to the non-linearities implicit
in the $u$ dependence of
$f_p(\vec r, u)$.  Nevertheless a lot of the
qualitative behavior can be guessed.

If the driving force is sufficiently small,
then it will  be insufficient
to overcome the pinning forces.  But if $F$ is
increased slowly, it may overcome the pinning of
some small segment of the interface which can
then jump forwards only to be stopped by
stronger pinning forces or by the elastic forces
from neighboring still-pinned parts of the
interface.  But if the drive is larger, the
neighboring regions may themselves not be
strongly enough pinned to resist the increase in
stress from the jumping section and may
themselves jump forward leading to an
``avalanche'' of some larger region of the
interface; this process might or might not
eventually stop \cite{BTW}.  If the force is large
enough---and certainly if it exceeds the maximum
$f_p$---then it is not possible for the interface
to be pinned and the interface will move forward
with some average velocity $\overline v$, albeit
via very jerky motion  both in space and time.

In addition to the  basic questions raised in
the Introduction, we will be interested, in the
regime where avalanches can stop, in the
statistical properties of the sizes and dynamics
of the avalanches \cite{BTW}\cite{MidN}\cite{Int2}.

The existence of a {\it unique critical force} can be
established by a simple convexity argument that
is valid if $J(\vec r,t)$ is non-negative \cite{NoP}. 
If two configurations
$u_a$ and
$u_b$ of the interface at time $t_0$ have the
property that one is ``ahead'' of the other,
i.e. $u_a (\vec r, t_0) > u_b (\vec r, t_0)$
for all $\vec r$; then $u_a$ will be ahead of
$u_b$ at all later times.  This can be seen by
assuming the contrary and then considering the
putative first time $t_1>t_0$ at which there is a
point of contact; say $u_a(\vec r_1,
t_1)=u_b(\vec r_1, t_1)$ at some $\vec r_1$.  Then
\begin{eqnarray}
\frac {\partial}{\partial
t}\left[u_a(\vec r_1, t)-u_b(\vec r_1, t) 
\right]|_{t=t_1}
=\sigma[\vec r_1, \{u_a\}]-\sigma[\vec r_1,
\{u_b\}]\nonumber\\
=\int d \vec r'\int^{t_1}dt'\ J(\vec
r_1-\vec r', t_1-t') [u_a(\vec r', t')-u_b(\vec
r', t')]>0
\label{eq:B5}
\end{eqnarray}
since the pinning force at $\vec r_1$ is the
same in both configurations and therefore cancels out. 
By assumption the last expression in
Eq~(\ref{eq:B5}) is positive as long as $J$ is
non-negative so that for
$t>t_1$,
$u_a$ is again ahead of $u_b$ violating the
assumption.

The condition that 
\begin{equation}
J(\vec r,t)\geq0
\end{equation}
 for all
$\vec r,t$ plays an important role  in the
theoretical analysis and 
frequently also in the physics of these  types
of systems. We will refer to models with this
convexity property as {\it monotonic}; they have the
property that if the displacements and the
driving force $F(t)$ increase monotonically with
time, then so will the total ``pulling
force''---see later---on any segment. Except in
the final section we will {\it focus solely on
monotonic models}.

We have shown that in monotonic models one
configuration that is initially behind cannot
``pass'' another that is ahead of it \cite{NoP};
therefore stationary and  continually moving solutions
{\it cannot coexist} at the same
$F$; therefore {\it $F_c$ is unique}. This is a
big simplification and one that will not occur
generally, in particular not in some of the
systems that we  discuss in the last section.

For forces well above $F_c$, one can use
perturbative methods to study the effects of the
random pinning and compute, for example: the
mean velocity,
$\overline v (F)$, the spatio-temporal
correlations of the local velocities, and
responses to additional applied forces \cite{CDWF}. 
Near
$F_c$, however, life is much more complicated as
is usually the case near critical
points---but for conventional equilibrium
critical points the theoretical framework for
dealing with the complexities is well
established \cite{HohH}.  The most interesting behavior
occurs in the critical regime; in particular one
might expect processes---such as avalanches---to
occur on a wide range of length and time scales. 
In order to make real progress, we will have
to---at least initially---make further
simplifications or approximations.  

One of the
lessons from equilibrium critical phenomena is
that analyzing simple models exactly or by
controlled approximations is more useful than
analyzing more realistic models by uncontrolled
approximations; we will thus take the former
route.  But we must first find some clues as to
what simplifications  preserve---what
we hope will be---the most essential features.

Near the critical force, at any given time most
of the interface will be moving very slowly if
at all so that the left hand side of
Eq~(\ref{eq:B2}) will be close to zero.  Thus a
first try might be to replace the actual dynamics
with the {\it adiabatic} approximation that the
forces at every point always balance exactly. 
Let us focus on one point $\vec r$ on the
interface and divide the stress $\sigma(\vec
r)$, Eq~(\ref{eq:B3}), into the local part
$\tilde{J}u(\vec r,t)$ and the non-local part,
$f_\sigma,$ involving $u(\vec r', t')$ for
$r'\not= r$ with
\begin{equation}
\tilde{J}\equiv\int
d\vec r'\int dt' J(\vec r', t').
\label{eq:B6}
\end{equation}
    We then have
a balance between the local 
force, $ f_p+\tilde{J}u$ and the {\it pulling
force}, 
\begin{equation}
\phi (\vec r,t) \equiv
f_\sigma (\vec r,t) + F.
\label{eq:B7}
\end{equation}
The adiabatic approximation corresponds to
\begin{equation}
f_p(\vec r,t)\approx \phi(\vec r, t) -\tilde{J}u
(\vec r,t).
\label{eq:B7a}
\end{equation}
But generally, because of the non-linearities in
$f_p(u)$, for a fixed $\phi$ ``applied'' to
$u(\vec r)$ there can be multiple values of $u$
which satisfy Eq~(\ref{eq:B7a}); as we shall see
these play an important role in the physics.

At this point, it is helpful to be more
concrete. Let us consider a simple model of the
pinning consisting of pinning sites
$u^p_{\alpha}(\vec r)$ distributed for fixed
$\vec r$ with random spacings between the
$u^p_{\alpha} (\vec r)$,
\begin{equation}
\Upsilon_\alpha(\vec r)\equiv u^p_{\alpha+1} (\vec
r)-u^p_{\alpha}(\vec r)
\label{eq:B8}
\end{equation}
 drawn, for each $\vec r$, independently from a
distribution
$\Pi(\Upsilon) d\Upsilon$. The pinning force
$f_p[(\vec r,u(\vec r)]=0$ except if
$u(\vec r)$ is equal to one of the pinning
positions, while for
$u(\vec r)=u^p_\alpha(\vec r)$, $f_p$ can take
any value between zero and a {\it yield strength},
$f_y$, which is the same for each pin.  A typical
realization of the pinning force
$f_p(u)$ on some segment of the interface is
plotted in Fig.~2a.  Note that for a {\it fixed
$\phi$}, there are {\it several possible values of $u$
} given by the intersection of the line
$\phi-\tilde{J}u$ with  
$f_p(u)$. 

\begin{figure}
  \begin{center}
   \leavevmode
   \epsfxsize=2.5truein
   \epsfbox{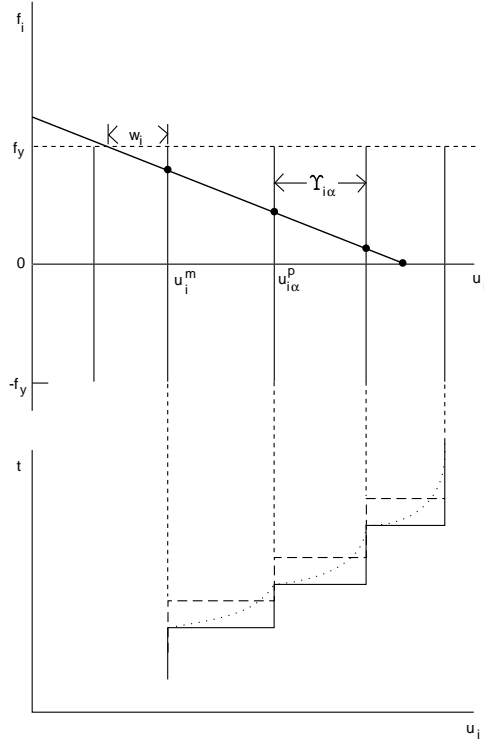}
  \end{center}
\caption{a) Simple model of the forces on
one segment of an interface. The segment can be pinned
at the positions, $u^p_{i\alpha}$, of the vertical lines at which
the pinning force can take any value up to the yield
strength $f_y$. The
intersections of the ``comb'' representing the pinning
force $f_p(u_i)$ and the diagonal line
$\varphi-\tilde{J}u_i$ with $\varphi$ the total pulling
force from the applied force and other segments of the interface, are the 
possible stationary positions of $u_i$
indicated by the dots. The one of these with the
smallest $u_i$, $u_i^m$, plays a special role as
discussed in the text. The amount $\Delta\varphi$
that $\phi$ needs to increase by to depin the
segment from this pinning position is $w_i\tilde{J}$.
b) Dynamics of the same segment of the interface as
the pulling force is increased. The actual $u_i(t)$ (dotted),
the adiabatic approximation to this (solid), and the time
delayed approximation (dashed) that is used in the analysis in
the text are all shown.}
  \label{fig:2}
\end{figure}

If $\phi$ is increased, then the
particular (history dependent) force-balanced
position $u(\phi)$ that the interface point is
following adiabatically can become unstable---for
example, the configuration denoted by the circle
in Fig.~2a---and $u$ must jump to a new
position.  During the jump
$\eta\frac{\partial u}{\partial t}$ is clearly
{\it not} small and Eq~(\ref{eq:B7a}) will not be
satisfied.  But if such jumps occur more rapidly
than the time scales of interest then their
primary effect will be a time lag between the
actual
$u[\phi(t)]$ and an adiabatic solution 
$u_{ad}[\phi(t)]$ to Eq~(\ref{eq:B2}).  A way to
capture this feature while preserving both the
physics of the time delays and the conceptually
simplifying separation of motion into adiabatic
and jump parts (with $\frac{\partial
u}{\partial t}=0$ in the adiabatic parts for
the pinning model illustrated in Fig~\ref{fig:2}),
is to require 
\begin{equation}
u[\phi(\vec r,t)] = u_{ad}[\phi(\vec r,t
-t_d)]
\label{eq:B9}
\end{equation}
 with some fixed (microscopic) delay time
$t_d$.  This is illustrated in Fig~\ref{fig:2}a. Note 
that, formally, this can be accomplished, by taking
$\eta\to 0$ and
$J(\vec r, t)= J(\vec r)\delta(t-t_d)$.

\subsection*{III. Infinite-range model: mean
field theory}

The above discussion in terms of the local
pulling force $\phi(\vec r, t)$ suggests that we
could try to analyze the system crudely by
assuming that the spatial and temporal
fluctuations in $\phi(\vec r,t)$ are small so that
$\phi$ can be replaced by some sort of time
dependent average $\overline\phi(t)$ which would
then need to be determined self-consistently from
the behavior of the neighboring regions that
contribute to the stress at our chosen point
$\vec r$ \cite{CDWF}.  This is very analogous to the
well known {\it mean field} approximation to
conventional phase transitions: for example, in a
magnetic system  the statistical mechanics (or
dynamics) of a  single spin $S(\vec r)$ in the
presence of a local effective field,
$h_{eff}(\vec r)$ from its neighbors---the mean
field---is analyzed and the mean field determined
self-consistently from the condition that at all
points the {\it assumed} $<\!S\!>$ entering the mean
field
$h_{eff} (\vec r)=\sum_{
\vec r'} J(\vec r-\vec r') <\!S (\vec r')\!>$, is
the same as the {\it computed} $<\!S\!>$.  One of
the advantages of this approximation is that it
has a well-defined regime of validity: in the
limit that the range of the interaction is very
long [or more properly that the effective number
of neighbors, $\left(\sum_{\vec
r}|J(\vec r) |\right)^2/\sum_{\vec r} |J(\vec
r)|^2$ is large] then the mean field theory
becomes exact.  But---users beware!---for fixed,
but finite range interactions it can still fail
near critical points as we shall see.

In order to obtain analytic results in at least
{\it some} model, we will study a strictly mean
field limit where each discrete segment of the
interface---we label simply by a subscript $i$ since
it is no longer really a spatial
coordinate---has an independent random pinning
force $f_p^i(u_i)$,  of the form in Fig~\ref{fig:2},
and the $N$ segments are {\it all} coupled
together by a uniform coupling
\begin{equation}
J_{ij}=\frac{J(t)}{N}
\label{eq:B10}
\end{equation}
i.e., {\it infinite range } forces.
(Note that Eq~(\ref{eq:B10}) includes a
self-coupling piece but its effects are
negligible in the desired
$N\to\infty$ limit.)  Much can be done for
general non-negative $J(t)$ and more complicated
forms of $f_p(u)$ using the actual dynamical
evolution Eq~(\ref{eq:B2}) \cite{CDWNF}, but to keep things
simple we will use the time-delayed adiabatic
approximation discussed above with
\begin{equation}
J(t)=\tilde{J}\delta (t-t_d);
\label{eq:B11}
\end{equation}
 and the form of
$f_p^i$ of Fig~\ref{fig:2}a with independent randomness
for each $i$. For simplicity, we will focus on the
strong pinning limit which corresponds to
\begin{equation}
f_y>\tilde{J}\Upsilon_{max}.
\end{equation}
  It is left to the
reader to show that including some of the more
``realistic'' features within the infinite range model
does not change the qualitative or other universal
aspects of the results.

Our task is now simple, at least in principle:
we assume some mean field 
\begin{equation}
{\overline\phi }(t)=F+\tilde{J}\frac{1}{N}
\sum_i u_i (t-t_d),
\label{eq:B12}
\end{equation}
compute the evolution of each $u_i(t)$ (from
$\phi_i=\overline\phi$ for all $i$) from some
set of initial conditions, and then adjust our
guessed ${\overline\phi}(t)$ until the computed
\begin{equation}
<u(t)>\equiv\frac{1}{N}\sum_i u_i(t)
\label{eq:B13}
\end{equation}
 is equal
to
$\left[{\overline\phi}
(t+t_d)-F\right]/\tilde{J}$ for all $t$. 

We
first try the  simplest possibility: a
{\it constant} ${\overline\phi}(t)$.  We can
proceed graphically.  From Fig~\ref{fig:2}a we
select for each $i$ one of the possible
stationary values of $u_i$.  We have many
choices; the only constraint being that
\begin{equation}
<u>=({\overline\phi}-F)/\tilde{J}.
\label{eq:B13a}
\end{equation}
 But we
must be careful: If we start choosing too many
large $u_i$'s, we may find that $<u>$ will
become too large.  We can thus ask: what are the
minimum and maximum possible $<u>$ for a given
$\overline\phi$?  The minimum will turn out to
be of primary interest so we focus on this: for
each $i$, the minimum $u_i$,
$u_i^m(\overline\phi)$ corresponds to the first
pinning position---i.e. one of the
$\{u^p_{i\alpha}\}$---to the right of the intersection
of the line $f=\phi-\tilde{J}u$ with the line
$f=f_y$ that passes through the tips of the
``comb'' --representing the yield strength-- in Fig.~2a\footnote{The
strong 
pinning
condition $f_y>\tilde{J}\Upsilon_{max}$ ensures that
$u_i^m$ is at a pinning position. The  general
case can be worked out similarly}.  Since the peaks are
randomly positioned,
\begin{equation}
<u_i>^{min}=<u_i^m(\overline\phi)>=
(\overline\phi-f_y)/\tilde{J}+<w_i>
\label{eq:B14}
\end{equation}
 with $w_i$
the distance to the next pin which has the
distribution\footnote{We use notations like ``${\rm
Prob}(w)$'' to mean the probability that the
continuous variable $w$ is in the range $w$ to
$w+dw$, divided by $dw$; i.e. ${\rm Prob}(w)$ is the probability
{\it density}  (usually called by physicists
``distribution'') of $w$. One must remember, however, that
if variables are changed e.g. from $w$ to $w'$, then
there is a Jacobian needed: ${\rm Prob}(w')=\left(\frac{dw}{dw'}
\right) {\rm Prob}(w)$. }:
\begin{equation}
{\rm Prob}(w)=
\int^\infty_w \frac{1}{\Upsilon}\left[
\frac{\Upsilon \Pi(\Upsilon) d\Upsilon}
{\overline\Upsilon}
\right].
\label{eq:B16}
\end{equation}
Here the quantity in parentheses is the
probability distribution that a random point  is
in an interval of width $\Upsilon$ between pins;
this includes the factor of
$\Upsilon/\overline\Upsilon$ with
\begin{equation}
\overline\Upsilon
\equiv\int^\infty_0\Upsilon \Pi(\Upsilon)
d\Upsilon
\end{equation}
 because of the presence of more points in
 wider intervals.  Integration of
Eq~(\ref{eq:B16}) by parts yields
$<w>= \overline{\Upsilon^2}/(2\overline
\Upsilon)$ so that 
\begin{equation}
<u_i>^{min}=(\overline\phi-f_y)/\tilde{J}+
\frac{\overline{\Upsilon^2}}{2\overline
\Upsilon}=\frac{F}{\tilde{J}}+<u_i>-
\frac{f_y}{\tilde {J}} +
\frac{\overline{\Upsilon^2}}{2\overline
\Upsilon}
\label{eq:B17}
\end{equation}
from Eqs~(\ref{eq:B12}) and (\ref{eq:B14}).  For
self consistency, we must therefore have
\begin{equation}
F\leq
F_c=f_y-\tilde{J}\frac{\overline{\Upsilon^2}}
{2\overline\Upsilon}
\label{eq:B18}
\end{equation}
a non-trivial result for the {\it critical force}
above which no static solutions are possible.
Note that as the interaction strength, $\tilde{J}$,
is increased, the critical  force {\it decreases}.
Physically, this is a consequence of the elasticity
causing the system to average over the randomness
more effectively: pulling a stiff object over a rough
surface is easier than pulling a flexible one.

For $F<F_c$ the
number of stable solutions, $N_s$ will be
exponentially large with an ``entropy'' per
segment $\frac{\ln N_s}{N}$ which is of
order one well below $F_c$ but decreases to zero
at
$F_c$ as most of the $u_i$'s will then  need to
take their minimum values to ensure
self-consistency.

What happens as $F$ is increased slowly from
below $F_c$ so that the non-adiabaticity is
negligible? Since this will certainly result in
$<u>$ and hence $\overline\phi$ increasing, we
can understand the behavior from Fig~\ref{fig:2}.  As
$\overline\phi$ increases, some segments become
unstable and jump to their next pinning
positions.  But this cannot happen unless they
are stuck on the pin with smallest
$u^p_{i\alpha}$, i.e.
$u_i^m(\overline\phi)$.  Furthermore, after a
jump $u$ will again be on the new smallest
$u_{i\alpha}^p$ for the increased
$\overline\phi$.  Thus the $u_i$'s are gradually
swept   to their minimum stable positions
as $F$ is increased towards $F_c$.

Above $F_c$, the segments continue to jump from
one $u_i^m(\overline\phi)$ to the next.  But
now the time delays must play a role.  If we
assume a solution which progresses uniformly on
average, $<u>=vt$, then 
\begin{equation}
\overline\phi=\tilde
{J}vt-\tilde{J}vt_d+F.
\label{eq:B19}
\end{equation}
[Note that to ensure that $u$ does not stop
between pins, we again need the strong pinning
condition
$\tilde{J}\Upsilon_{max}<f_y$]
  With all $u_i=u^m_i(
\overline\phi)$, our earlier analysis
immediately yields self-consistency only when
\begin{equation}
\overline{v}=\frac{F-F_c}{\tilde{J}t_d}
\label{eq:B20}
\end{equation}
so that near criticality
\begin{equation}
\overline{v}\sim(F-F_c)^\beta
\label{eq:B21}
\end{equation}
with the  {\it critical exponent} 
\begin{equation}
\beta=\beta_{MF}=1
\label{eq:B22}
\end{equation}
in this infinite range mean-field model \cite{MFB}.

\begin{figure}
  \begin{center}
    \leavevmode
    \epsfxsize=3.5truein
    \epsfbox{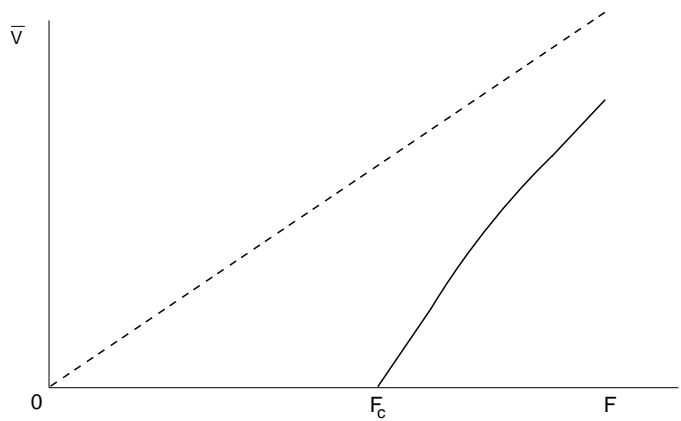}
  \end{center}\caption{Velocity versus driving force in a typical mean
field model is indicated by the solid line. Note the
linear dependence of $v$ on $F$ just above $F_c$. The
dashed line is the behavior in the absence of pinning.}
  \label{fig:3a}
\end{figure}

Note that a comparison of Eq~(\ref{eq:B20}) for
$F>>F_c$ and the original dynamic equation
(\ref{eq:B2}), suggests that a natural choice is
$t_d=\frac{\eta}{\tilde{J}}$ so that
$\overline{v}=\frac{F}{\eta}$ for large $F$.  The
actual non-adiabatic processes can  be seen
to give rise to an effective $t_d$ of roughly this
magnitude.  But the breakdown of the jump
approximation for large $F$ will make
$\overline{v}$ {\it not} strictly linear for $F>F_c$.
Nevertheless near
$F_c$, the mean velocity will still be
characterized by the exponent $\beta=1$. A
typical mean-field $\overline{v} (F)$ curve is
shown in Fig~\ref{fig:3a}.

\begin{figure}
  \begin{center}
    \leavevmode
    \epsfxsize=3.5truein
    \epsfbox{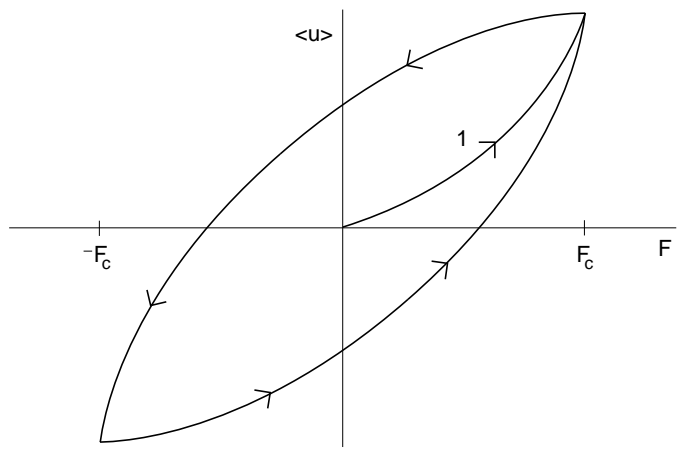}
  \end{center}
\caption{Schematic of hysteresis loops that occur as the
force is increased from zero to the critical force,
decreased  to the critical force in the opposite
direction, and then cycled between these values. The
direction of change of $F$ is indicated by the arrows
with the ``1'' denoting the first increase.}
  \label{fig:3b}
\end{figure}

In addition to the critical force and the steady
state velocity, in our simple mean field 
model one can compute other ``macroscopic''
properties e.g., hysteresis loops as $F$ is
increased and decreased---sketched in Fig~\ref{fig:3b}
\cite{MidN}---and responses to a time dependent
additional driving force which in the pinned phase
will depend, due to the metastability, on the past
history,  the sign of the perturbations,  etc
\cite{CDWF}.  Some of the properties---like the
exponent
$\beta$---will be universal within a broad class
of  mean field models, while others will depend 
on details.  Nevertheless, substantial qualitative
insight can be obtained that gives useful clues
to the behavior of finite---in particular
real two-dimensional---interfaces.  Furthermore,
experience with equilibrium critical phenomena
suggests that some of the asymptotic {\it
forms} near
$F_c$, such as Eq~(\ref{eq:B20}) will be correct
in realistic models if the dimension of the
``interface'' is sufficiently large or the
interactions are sufficiently long range
\cite{Kar}\cite{EK}.  We will return to these
questions later, but for now we stick to the simple
mean field model and ask what can be learned about
``microscopic'' properties, in particular the
properties of avalanches that occur as the driving
force is increased towards $F_c$.

\subsection*{IV. Avalanche statistics and
dynamics}

In the mean field model introduced in the
previous section, the statistics and other
properties of the avalanches of jumps that
occur as the driving force is increased slowly can
be worked out in substantial detail
\cite{AvPer}\cite{Dah}.  We will carry out the
analysis using methods which can be generalized to
provide useful information about the behavior with
more realistic interactions.

Let us consider what happens for $F<F_c$ when
$F$ is increased by a very small amount.  If the
increase is sufficiently small, then no segments
will jump.  But a slightly bigger
increase---typically of order
$\frac{1}{N}$---will result in one jump.  Let us
call the time of this jump $t=0$ and measure times
in units of the delay time $t_d$ so that with $F$
 held fixed after the avalanche starts, we have
simple discrete time dynamics.  The first jump can
trigger $n_1$ other jumps at time $t=1$, with
these triggering
$n_2$ further ones at $t=2$ etc. As long as the total
number of jumps is finite, then in a large system the
mean
$<u>$, and hence
$\overline\phi$ will have only advanced by
an amount of order $\frac{1}{N}$.  Thus
all that will matter is the distribution of
segments which are very close to jumping, i.e.
those which will jump when $\overline\phi$ is
increased by a small amount $\Delta\varphi_i$.
>From Fig.~2a, we see that
$\Delta\varphi_i=\tilde{J}w_i$.  For large $N$, all but
very special ways of preparing the conditions before
the avalanche starts will yield a distribution of
these small $\Delta\varphi_i$ which are independent
and randomly distributed with (initial-condition
dependent) density
\begin{equation}
\rho\equiv\rho(\Delta\varphi_i=0);
\label{eq:C2}
\end{equation}
 $\rho$ thus
measures a {\it local susceptibiltiy to jumping}.  We
can now immediately conclude something about the
mean number of jumps $<n_t>$ at a time
$t$ after the initial jump. Since the
$n_{t-1}$ jumps at  time $t-1$ will
cause an increase in $\overline\phi$ by, on
average, $\tilde{J}<\Upsilon> n_{t-1}$
(where we have used $<\Upsilon>$ rather than
$\overline{\Upsilon}$ since the distribution of
$\Upsilon$'s for the almost unstable segments and
hence
$<\Upsilon>$ could depend on the initial
conditions).  This will cause, on average,
$\rho<\Upsilon> n_{t-1}$ jumps at time $t$,
i.e.
\begin{equation}
<n_t>=\rho<\Upsilon>\tilde{J}
<n_{t-1}>.
\label{eq:C3}
\end{equation}

The crucial parameter is thus $\rho<\Upsilon>
\tilde{J}$; if this is greater than one the
avalanche will runaway. 
If the system is below $F_c$ as we have assumed,
it will eventually  be stopped only when a finite
fraction of the segments have jumped and the system
has found a stable---and more
typical---configuration.  But if
$\rho <\Upsilon>
\tilde{J}<1$, then the expected total size of an
avalanche 
\begin{equation}
s\equiv\sum_i\Delta u_i
\end{equation}
is simply 
\begin{equation}
<s>=\frac{<\Upsilon>}{1-\rho<\Upsilon>\tilde{J}}.
\label{eq:C4}
\end{equation}
As we shall see however, this is {\it not} the typical
size: even very close to criticality; i.e.
\begin{equation}
\epsilon\equiv 1-\rho<\Upsilon>\tilde{J}<<1,
\label{eq:C4a}
\end{equation}
most avalanches will be small.

The distribution of avalanche sizes, as well as
other interesting information on their dynamics
etc., can be obtained from generating function
methods.  Since these are a widely applicable
tool, we will go through some of the details.  For
simplicity we work in
units with
\begin{equation}
<\Upsilon>=1
\label{eq:C4b}
\end{equation}
 and 
\begin{equation}
\tilde{J}=1.
\label{eq:C4c}
\end{equation}
We are interested in the time evolution of the
displacements, in particular the increments
\begin{eqnarray}
m_t&=&\sum_i[u_i(t)-u_i(t-1)]\label{eq:C4d}\\
&=&\sum^{n_t}_{\alpha=1}\Upsilon_{t\alpha}\nonumber
\end{eqnarray}
where the $\{\Upsilon_{t\alpha}\}$ are the magnitudes
of the $n_t$ jumps that occur at time $t$ (with
$m_t=0$ if $n_t=0$). 
The total size is simply
\begin{equation}
s=\sum^\infty_{t=0}m_t.
\label{eq:C4e}
\end{equation}
The joint probability
distribution of all the $\{m_t\}$ given the
initial jump $n_0=1$
\begin{equation}
{P}\{m_t\}\equiv{\rm Prob}[m_0, m_1, m_2
\dots|n_0=1],
\label{eq:C5}
\end{equation}
contains the information of interest. Note that
vertical bars as in Eq~(\ref{eq:C5})  denote
``given''; i.e. conditional probability. It is
useful to define a {\it generating function} of the
distribution including  all times up to $T$
\begin{equation}
\Gamma_T\{\mu_t\}\equiv<exp(i\sum^T_{t=0}\mu_t
m_t)>_P\ ;
\label{eq:C6}
\end{equation}
(usually we will drop the $P$). Note that $\Gamma_T$ is simply the Fourier 
transform of $P$ restricted to times $\leq T$. We can 
derive a recursion relation for
$\Gamma_{T }$ in terms of $\Gamma_{T-1}$. 
For a given $m_{T-1}$, the number of jumps
triggered at time $T$ will be Poisson
distributed with mean $\rho m_{T-1}$ i.e.
\begin{equation}
{\rm Prob}\left(  n_T|  m_{T-1} \right)=
\frac{e^{-\rho m_{T-1}}}{n_T!}(\rho
m_{T-1})^{n_T}.
\label{eq:C7}
\end{equation}
Then, since $m_T$ depends only on $m_{T-1}$,
we can compute
\begin{eqnarray}
& &<e^{i\mu_T m_T}|\{m_t\}_{t<T}>= <e^{i\mu_t
m_T}|m_{T-1}>\label{eq:C8}\\
&=&\sum^{\infty}_{n_T=0}\left\{{\rm
Prob}(n_T\mid m_{T-1})\prod^{n_T}_{\alpha=1} 
\left[ \int d\Upsilon_{T\alpha}\Pi
\left(\Upsilon_{T\alpha} 
\right) e^{i \mu_T\Upsilon_{T\alpha}}
\right]\right\}\nonumber\\
&=&\exp \left[ \rho m_{T-1} \left(
<e^{i\mu_T\Upsilon}>-1
\right)
\right]\nonumber
\end{eqnarray}
The last equality follows from Eq.~(\ref{eq:C7}); the resulting expression
has similar $m_{T-1}$ dependence to the
$e^{i\mu_{T-1}m_{T-1}}$ factor in $\Gamma_{T-1}$.

We thus find that 
\begin{equation}
\Gamma_T[\mu_0, \dots, \mu_T] = \Gamma_{T-1}
[\mu_0, \dots, \mu_{T-2}, \lambda_{T-1}]
\label{eq:C9}
\end{equation}
with
\begin{equation}
\lambda_{T-1}=\mu_{T-1}-i\rho(<e^{i\lambda_
T\Upsilon}>-1)
\label{eq:C10}
\end{equation}
where
\begin{equation}
\lambda_T=\mu_T.
\label{eq:C11}
\end{equation}
We can now iterate with the recursion relation
Eq~(\ref{eq:C10}) from the ``initial'' condition
Eq~(\ref{eq:C11}), eventually obtaining 
\begin{equation}
\Gamma_T\{\mu_t\}=\Gamma_0\left(\lambda_0\{\mu_t\}\right)  
=<e^{i\lambda_0\{\mu_t\}\Upsilon}>
\label{eq:C12}
\end{equation}
(since $n_0=1$).
All the information has thus gone into
$\lambda_0$.  As long as the system is stable,
i.e. $\rho\leq1$, we can simply take
$T\to\infty$ to recover the full information.
[If $\rho>1$, then there is a non-zero (and
computable) possibility that $s=\infty$,  and
more care is needed.]

To get the probability distribution of $s$, we
simply set all $\mu_t=\mu$ and then 
\begin{equation}
{\rm Prob}(s)=\int_\mu e^{-i\mu s} e ^{i\lambda^
*(\mu)}
\label{eq:C13}
\end{equation}
with 
\begin{equation}
\int_\mu\equiv\frac{1}{2\pi}\int^{\infty}_{-\infty}
d\mu
\label{eq:C14}
\end{equation}
and $\lambda^*(\mu)$  the (stable) {\it fixed
point} solution to Eq~(\ref{eq:C10}). Let us first
consider the behavior for large $s$.  This will be
dominated by the singularity in the lower half
complex $\mu$ plane nearest to the real
axis---a general property of Fourier transforms
that can be seen by deforming the integration
contour away from the real axis until it
encounters a singularity.  We only expect to have
large
 avalanches for
\begin{equation}
\epsilon\equiv1-\rho
\label{eq:C15}
\end{equation}
small, so the interesting regime is $\mu$ and
$\epsilon$ small which suggests $\lambda^*$ small. 
We find that to leading  order,
\begin{equation}
\lambda^*\approx\frac{\left[-i\epsilon+
\sqrt{-\epsilon^2+2i b \mu}\right]}{b}
\label{eq:C16}
\end{equation}
with
\begin{equation}
b\equiv<\Upsilon^2>
\label{eq:C16a}
\end{equation}
and the sign of the square root that has
positive imaginary part for $\mu$ real 
chosen.  For $\mu\to0$, this gives $\lambda^*=0$
as it must for normalization of probability
$\Gamma\{\mu_t=0\}=1$. The integration contour in
Eq~(\ref{eq:C13}) can be deformed so that it is
dominated by the cut at
$\mu=-\frac{1}{2}i\epsilon^2/b$ for large $s$ and
we thus have, after replacing the dummy variable
$\mu$ by
$\mu-\frac{1}{2}i\epsilon^2/b$ and expanding in small
$\mu$,
\begin{equation}
{\rm Prob}(s)\approx\int_{\mu} i\sqrt {2i\mu/b}\;e
^{-i\mu s}e^{-s\epsilon^2/(2b)}
\label{eq:C17}
\end{equation}

By ``power counting'', we see that the branch
cut must yield a $\frac{1}{s^{\frac{3}{2}}}$
dependence; hence for large $s$,
\begin{equation}
{\rm Prob} (s)\sim
\frac{e^{-s\epsilon^2/(2b)}}{s^{\frac{3}{2}}}
\label{eq:C18}
\end{equation}
\cite{AvPer}. Note that for small $\epsilon$, the mean
$<s>$ {\it is dominated by   large $s$ avalanches}
which are {\it rare} since  
\begin{equation}
{\rm Prob}(s\sim\frac{1}{\epsilon^2})\sim {\rm Prob}(s>\frac{1}{\epsilon^2})
\sim\epsilon.
\label{eq:C19}
\end{equation}
We see from Eq~(\ref{eq:C18}) that these yield
$<s>\sim\frac{1}{\epsilon}$ as expected.

Two questions now arise: First, can we trust
this heuristic calculation? And, second, is the
large $s$, small $\epsilon$ behavior  in
Eq~(\ref{eq:C18}) generic?  The latter we have 
answered already: the  large
$s$ behavior is the same, (except for the coefficient
$b$) as long as
$b=<\Upsilon^2><\infty$. [The reader is
encouraged to find the behavior associated with
a power law tail in the distribution of 
\renewcommand{\thefootnote}{\fnsymbol{footnote}}
\setcounter{footnote}{1}]
$\Upsilon$\footnote{Note that arbitrarily large
jumps can only occur in models that also have
arbitrarily large local yield stresses, $f_y$.}.

As far as justifying the result
Eq~(\ref{eq:C18}), for the case in which all the
jumps are the {\it same}, $\Upsilon=1$,  one can
compute ${\rm Prob}(s)$ {\it exactly}  by changing the
variable of integration to
$z=e^{i\lambda^*(\mu)}$ so that the integral
in Eq~(\ref{eq:C13}) circles  an $s$th order pole
at
$z=0$. Cauchy's theorem then yields
\begin{equation}
{\rm Prob}(s)=e^{-\rho s}\frac{(\rho
s)^{s-1}}{s!},
\label{eq:C20}
\end{equation}
with, of course, $P(s=0)=0$.
From the limiting large $s$ form 
\begin{equation}
s!\approx s^s
e^{-s}\sqrt{2\pi s}
\label{eq:C20a}
\end{equation}
 the asymptotic behavior 
Eq~(\ref{eq:C18}) is found confirming the validity
of the approximations made in our first
derivation of this result. Note that the
$\rho$ dependence is simply via the
$\frac{1}{\rho}(\rho e^{-\rho})^s$ factor in
Eq~(\ref{eq:C20}). It is nice that the exact
result can be found in this case, but in general,
asymptotic methods like those we have used above give
more understanding and are more widely
applicable.  Nevertheless, to convince skeptical
colleagues, a few exact results are useful!

In addition to the distribution of avalanche
sizes, we are also interested in their temporal
evolution.  For example,  one might ask what
is $<m_t|s>$, i.e. what is the time
development of an average event of size $s$? 
This can be computed using the generating
function. If we choose 
\begin{equation}
\mu_\tau=\mu
\mbox{  for }
 \tau\not=t
\label{eq:C20b}
\end{equation}
  and
\begin{equation}
\mu_t=\mu+\nu_t,
\label{eq:C20c}
\end{equation}
 then
\begin{equation}
\frac{1}{i}\left.\frac{\partial\Gamma_\infty}
{\partial\nu_t}\right|_{\nu_t=0}= <n_te^{i\mu
s}>
=\left.\frac{\partial\lambda_0}{\partial\nu_t}
\right|_{\nu_t=0}<e^{i\lambda^*(\mu)\Upsilon}>
\label{eq:C21}
\end{equation}
whose Fourier transform in $\mu$ yields 
\begin{equation}
\int m_t {\rm Prob}(m_t, s) dm_t=\int m_t Prob
(m_t|s) dm_t {\rm Prob}(s)=
<m_t|s>{\rm Prob}(s).
\label{eq:C22}
\end{equation}
But by multiple use of the chain rule and
Eq~(\ref{eq:C10}),
\begin{equation}
\frac{\partial\lambda_0}{\partial\nu_t}
=\left(\frac{\partial\lambda_t}
{\partial\mu_t}\right)_{\lambda_{t+1}}
\left(\frac{\partial\lambda_{t-1}}
{\partial\lambda_t}\right)_{\mu_{t-1}}
\times\dots\times
\left(\frac{\partial\lambda_0}
{\partial\lambda_1}\right)_{\mu_0}
\label{eq:C23}
\end{equation}
evaluated with all $\lambda_\tau=\lambda^*(\mu)$
and all $\mu_\tau=\mu$.

\begin{figure}[h]
 \begin{center}
 \leavevmode
    \epsfxsize=2.3truein
    \epsfbox{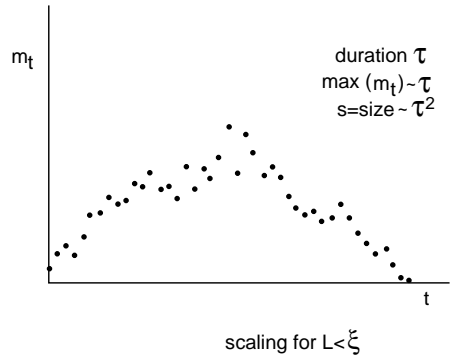}
  \end{center}
\caption{Typical avalanche in a mean field
model showing the increase in size of the avalanche,
$m_t$,  at each time step. Note that the fluctuations in
$m_t$ are larger when $m_t$ is larger.}
\label{fig:4}
\end{figure}

For the constant $\Upsilon=1$ case we obtain,
\begin{equation}
\left.\frac{\partial\lambda_0}{\partial\nu_t}
\right|_{v_t=0} =\left[\rho
e^{i\lambda^*(\mu)}\right]^t
\label{eq:C24}
\end{equation}
After shifting $\mu$ as in Eq~(\ref{eq:C17})  we see
that, for
$\epsilon$ small and $s$ and $t$ large, 
\begin{equation}
<m_t|s>\sim\frac{1}{Prob(s)}\int_{\mu} 
e^{-s\epsilon^2/2}
e^{-i\mu s+it\sqrt{2i\mu }}.
\label{eq:C25}
\end{equation}
This will be dominated by $\mu\sim\frac{1}{s}$
and hence the typical duration of an avalanche
$s$ will be 
\begin{equation}
\tau\sim\sqrt{s}.
\label{eq:C26}
\end{equation}
Note that this is much less then the maximum
possible duration $\tau_{max}=s-1$.
The integral in Eq~(\ref{eq:C25}) can be done
exactly (by writing $\mu=-i\frac{x^2}{2}$)
yielding
\begin{equation}
<m_t|s>\approx te^{\frac{-t^2}{2s}}
\label{eq:C27}
\end{equation}
for large $s$ {\it independent} of $\rho$.  Again,
the behavior for large $s$  and $1<<t<<s$ is
generic up to a coefficient $b$ that should appear as
in Eq.~(\ref{eq:C18}). For the particular constant
$\Upsilon$ case, the exact result can be computed from
Eq~(\ref{eq:C22}) yielding 
\begin{equation}
<m_t|s>=\frac{t+1}{(s-t-1)!}(s-1)!
s^{-t}\ \ \ {\mbox{ for }}\  0\leq t\leq s-1.
\label{eq:C28}
\end{equation}
Note that this result includes the rare tail of
events which have total duration $t$ of order $s$;
the asymptotic methods needed to obtain this are
a variant of those used above although terms
neglected in Eqs~(\ref{eq:C16},\ref{eq:C25})
become important in this region. A schematic of the evolution of a large
avalanche is shown in Fig~\ref{fig:4}.  Note that, on
average, an
avalanche which is going to be large starts with
$m_t$ growing linearly in time, {\it independent of
how big it will become}; this is very different
from the behavior of typical avalanches even near
to criticality  which are small. Nevertheless,
typical large avalanches will have large
fluctuations of $m_t$ away from
$<m_t|s>$; these can be studied by
computing, e.g. $<m_t^2|s>$ from which it can be
concluded that a
typical $m_t$ for  a large
$s$ avalanche  is the same order as $<m_t|s>$
except for $t>>\frac{1}{\sqrt{s}}$. The
probability that a large avalanche stops before
time $t$ can be computed from
${\rm Prob}(m_t=0|s)$ which is found from
$\Gamma_\infty\left(\mu_t\to +i\infty,\
\mu_\tau=\mu \mbox{ for } \tau\not=t\right)$.

So far, we have not attempted to relate the
local susceptibility to jumps, $\rho$, to the
original mean field model of the previous
section.  In general,
$\rho$ will depend on the past history. But on a
generic approach to $F_c$ from below (e.g. after
``training'' the system by a slow increase to
$F_c$ from
$F=-F_c$), $\rho$ will approach unity at $F_c$
and the cutoff 
$\tilde{s}\sim\frac{1}{(1-\rho)^2}$ in the jump
size distribution will diverge.  Right at $F_c$
there will be a power law distribution of
avalanches sizes; this is analogous to the power
law distribution of clusters that occur at the
critical point for conventional percolation
\cite{Per}.  

One might also hope  to make
connections to power law {\it spatial}
structures at percolation \cite{Per} as well as to
spatial correlations at conventional equilibrium
critical points.  But to do this we will
certainly have to move away from the
infinite-range mean field model.  This we do in
the next section.

\subsection*{V. Toy  model: spatial structure of
avalanches.}

The simple mean field model of avalanches
discussed in the previous section can be extended
in a relatively straightforward way to include
spatial and temporal structure like that which
arises for general non-negative stress transfer
functions
$J(r,t)$, i.e. monotonic models. Spatial
coordinates for 
$n(\vec r,t)$ are now needed and corresponding
generating function variables $\mu(\vec r, t)$.

In the mean field model, the probability that a segment $u(\vec r)$
jumps in a small time interval is proportional to
the increase in pulling force $\phi(\vec r)$ on it
in that interval  times a local jump
susceptibility
$\rho$. If we assume the same is true here, then we can generalize the recursion
relation equation Eq~(\ref{eq:C10}) to include
general
$J(\vec r, t)$: 
\begin{equation}
\lambda(\vec r,t)=\mu(\vec r,t)-i\rho
\int d\vec r'\int^\infty_t dt'J(\vec r'-\vec r,
t'-t)\{\exp[i\Upsilon\lambda(\vec r', t')]-1\}.
\label{eq:Da1}
\end{equation}
for the case with all jump displacements
$\Upsilon$  equal.
 Many quantities of interest are
computable by similar techniques to those in the
previous section, often in terms of the
spatio-temporal Fourier transform of $J(\vec
r,t),\ J(\vec q,\omega)$.

 The
mean number of jumps  at $\vec r$ at a time $t$
after an avalanche is triggered at $\vec r=0$ 
at time $t=0$ is
\begin{equation}
<m(\vec r, t)>=\int_{\vec q}\int_\omega
e^{-i\omega t}e^{i\vec q\cdot\vec
r}\frac{1}{1-\rho\Upsilon J(q,\omega)}.
\label{eq:D1}
\end{equation}
The critical point is thus still given by
\begin{equation}
( \rho\Upsilon\tilde{J})_{crit}=1
\label{eq:D1a}
\end{equation}
  with 
\begin{equation}
\tilde{J}=J(q=0,
\omega=0)
\label{eq:D2}
\end{equation}
 and the mean total size is
\begin{equation}
<s>=\frac{\Upsilon}{1-\rho\Upsilon\tilde{J}}
\label{eq:D3}
\end{equation}
as before. We will henceforth work in units with
$\Upsilon=\tilde{J}=1$.

A more interesting quantity is again the
conditional mean:
\begin{equation}
<n(\vec r, t)|s>=\frac{1}{Prob(s)}\int_\mu\int_{\vec q}\int_\omega
\frac{e^{-i\mu s} e^{i\lambda^*(\mu)} e^{i\vec q \cdot \vec r} e^{-i\omega t}}
{1-\rho\Upsilon J(q,\omega)e^{i\lambda^*(\mu)}}
\label{eq:D4}
\end{equation}
with $\lambda^*(\mu)$ the fixed point solution to
the mean field recursion relation
Eq~(\ref{eq:C10}) or, equivalently,
Eq~(\ref{eq:Da1}) with $\lambda$ and $\mu$
independent of $\vec r$ and $t$.  By changing
variables one can show that (as in the previous
section)  conditional statistics like
Eq~(\ref{eq:D4}) are independent of
$\rho$.  

The important physical quantity is
\begin{equation}
K(\vec q,w)\equiv1-J(\vec q,w)
\label{eq:D5}
\end{equation}
 which
embodies the information on the space and time
dependent elasticity.  Changing variables to
$\mu\to
\mu+\mbox{constant}$ and noting that for large
$s$, small $\mu$ will dominate as before, we
obtain
\begin{eqnarray}
<n(\vec q,\omega)|s>\approx\int_\mu\sqrt{2\pi s^3}
e^{-is\mu} \frac{1}
{-i\sqrt{2i\mu}+K(\vec q,\omega)}\nonumber
\\ =
\int d\lambda e^{-\frac{1}{2} s\lambda^2}\sqrt
{\frac{s^3}{2\pi}} \left[
\frac{\lambda^2}{\lambda^2+K^2(\vec
q,\omega)}\right]
\label{eq:D6}
\end{eqnarray}

For an interface with dissipative dynamics and
local elasticity, in the absence of pinning or
driving forces we have, after rescaling lengths and
times, simply
\begin{equation}
\frac{\partial u}{\partial t}=\nabla^2 u
\label{eq:D7}
\end{equation}
so that
\begin{equation}
K(\vec q,\omega)=-i\omega+q^2.
\label{eq:D8}
\end{equation}
But to understand the general behavior, and to
apply the results to other physical systems, we
would like to include the possibility of long
range elasticity, i.e.,
\begin{equation}
\int dt
J(\vec r,t)\sim\frac{1}{r^{d+\tilde{\alpha}}}
\label{eq:D9}
\end{equation}
in $d$-dimensions corresponding to the static 
\begin{eqnarray}
K_s(\vec q)\sim|q|^{\tilde{\alpha}}\equiv
K(\vec q,
\omega=0)\ \mbox{ if }\ 
\tilde{\alpha}<2 \nonumber\\
\mbox{ or }\  
K_s(\vec q)\sim q^2 \ \mbox{ if }\ 
\tilde{\alpha}>2.
\label{eq:D10}
\end{eqnarray} 
We thus consider the general
case of $K_s(\vec q)\sim|q|^\alpha$ with
$\alpha\leq2$ with
\begin{equation}
\alpha\equiv\mbox{min}(\tilde{\alpha},2).
\label{eq:D10a}
\end{equation}
 
 The
total displacement a distance $r$ from the
avalanche starting point, during an avalanche of
large size
$s$, is obtained from Eq~(\ref{eq:D6}) with
$\omega=0$. We must thus evaluate $\int_{\vec q}
e^{i\vec q\cdot \vec r} $ of the
last expression in Eq~(\ref{eq:D6}). For $r=0$,
all
$q$ can contribute but
$\lambda$ is small so that we can ignore the
$\lambda^2$ in the denominator, yielding 
\begin{equation}
<\Delta u(r=0)|s>\equiv \Upsilon<\int
dt\   n(r=0,t)|s>
\sim\int_{\vec q}\frac{1}{[K_s(q)]^2}
\label{eq:D11}
\end{equation}
which is of {\it order one independent of $s$} if
\begin{equation}
d>d_c(\alpha)=2\alpha.
\label{eq:D12}
\end{equation}
  We thus see the
appearance of a  special  {\it critical  
 dimension} above which no segment will jump more
than a few times even in an arbitrarily large
avalanche. Indeed, we will see that above the critical 
dimension driven interfaces will have only bounded small scale roughness.

For $d<d_c$, the integral in Eq~(\ref{eq:D11}) is
infinite so that small $q$ (i.e. small $K$)
dominates and more care is needed. The cutoff of
$\int_q$ of Eq~(\ref{eq:D6}) when $K\sim\lambda$
yields, with a typical
$\lambda\sim \frac{1}{\sqrt{s}}$ and hence
$q\sim s^{-\frac{1}{2\alpha}}$
\begin{equation}
<\Delta u(r=0)|s>\sim s^{1-\frac{d}{2\alpha}}.
\label{eq:D13}
\end{equation}
Note the appearance of a non-trivial exponent
relating $\Delta u$ and $s$. It depends, as is
usually the case for critical phenomena, on the
spatial dimension.  As mentioned in the
Introduction, Eq~(\ref{eq:D13}) is just the kind
of scaling law we expect near critical points.
Equation~(\ref{eq:D13}) relates characteristic
scales of displacement  to the characteristic
scales of avalanche size.  

We
can also say something about the spatial extent
and shape of large avalanches, by computing
$<\Delta u(r)|s>$. For
$d>d_c(\alpha)$, the integral in
Eq~(\ref{eq:D6}) will be cutoff for $q>r$ by the
$e^{i\vec q\cdot\vec r}$ oscillations and hence
dominated by $q\sim\frac{1}{r}$ yielding
\begin{equation}
<\Delta u(r)|s>\sim\frac{1}{r^{d-2\alpha}}
\label{eq:D14}
\end{equation}
for $1<<r<<s^{\frac{1}{2 \alpha}}$ where the
upper cutoff arises when
$K(q\sim\frac{1}{r})\sim\lambda\sim\frac
{1}{\sqrt{s}}$.

For $d<d_c$, on the other hand, as long as
$r<<s^{\frac{1}{2\alpha}}$, $\int_{\vec q}\frac
{e^{iq\cdot r}}{\lambda^2+K^2_s}$ for typical
$\lambda$ will be dominated by $q\sim
s^{-\frac{1}{2\alpha}}$ and 
\begin{equation}
<\Delta u(r)|s>\sim s^{1-\frac{d}{2\sigma}}\
\mbox{ for }\
r<<s^{\frac{1}{2\sigma}};
\label{eq:D15}
\end{equation}
i.e. the magnitude of  typical displacements is approximately independent of $r$ 
in this range.
 In all dimensions,
$\Delta u(r)$ will fall off rapidly for larger
$r$. The length
\begin{equation}
L\sim s^{\frac{1}{2\alpha}}
\label{eq:D16}
\end{equation}
is thus some measure of the {\it diameter} of an
avalanche.

Let us now try to interpret these results [Note
that the skeptic could compute e.g., $<[\Delta
u(r)]^2|s>$ etc. to provide further support for
the picture below].
For $d>d_c$, the fact that $<\Delta u(r)>$ is
much less than unity for $r>>1$ strongly
suggests that {\it most} segments will {\it not}
jump even if they are within $r<<L$ of the
avalanche center, rather only a fraction $\sim
1/r^{d-2\alpha}$ of them will jump, and these
typically only once or a few times.  The number
of sites that have jumped at all within a distance
$R<L$ of the origin is of order $R^{2\alpha}<<R^d$ so that the
avalanche is {\it fractal}.  The total number of
sites that jump, its ``area'' $A$ is thus, by
taking
$R\sim L$,
\begin{equation}
A\sim L^{d_f}\sim s<<L^d
\label{eq:D17}
\end{equation}
with the fractal dimension 
\begin{equation}
d_f=2\alpha\  \mbox{ for }\ 
 d>d_c.
\label{eq:D18}
\end{equation}

In lower dimensions, the picture is quite
different. The approximate independence of
$<\Delta u(r)|s>$ of $r$ for $r<<L$ suggests that
each site in this region jumps a comparable number
of times $\sim s^ {1-d/(2\alpha)}$  (with
fluctuations around this of the same order) and
hence the avalanche is {\it not} fractal but has
area
\begin{equation}
A\sim L^d\sim s^{\frac{d}{2\sigma}}
\label{eq:D19}
\end{equation}
while the typical displacement is
\begin{equation}
\Delta u(r)\sim L^\zeta
\label{eq:D20}
\end{equation}
for $r\leq L$ with 
\begin{equation}
\zeta=2\alpha-d.
\label{eq:D21}
\end{equation}
(for $d>d_c$, $\zeta=0$).

The distribution of $s$ at the critical point
that occurs at
$\rho\tilde{J}\Upsilon=1$, is the same as in the
mean field model.  This implies that 
\begin{equation}
\mbox{Prob (diameter}>L)\sim \frac{1}{L^\kappa}
\label{eq:D22}
\end{equation}
with
\begin{equation}
\kappa=\alpha.
\label{eq:D23}
\end{equation}
  The duration of an
avalanche with dissipative
dynamics---corresponding to
\begin{equation}
K(q,\omega)\approx-i\omega + |q|^\sigma 
\label{eq:D23a}
\end{equation}
---is given
simply by scaling, i.e., 
\begin{equation}
\tau\sim L^z\sim s^\frac{1}{2}
\label{eq:D24}
\end{equation}
with the {\it dynamical critical exponent}
\begin{equation}
z=\alpha.
\label{eq:D24a}
\end{equation}
Note that the  relation $\tau\sim\sqrt{s}$ is (not
surprisingly) the same as in the infinite-range
mean-field model.

We have found that in our toy model, many of the
properties of large avalanches near to the
critical point (actually {\it any} large
avalanche although they are rare away from
criticality), obey scaling laws which relate
various characteristic physical properties to
each other by power law relationships. For
example, for $d<d_c$, an avalanche of diameter
$L$ has typical size $s\sim L^{2\alpha/d}$,
displacement $\Delta u\sim L^\zeta$ and duration
$L^z$. This type of scaling behavior is one of
the key aspects of critical phenomena in both
equilibrium and non-equilibrium systems.  But
there is more: if we scale all lengths by a {\it
correlation length}
\begin{equation}
  \xi\sim\epsilon^{-1/\alpha}
\label{eq:D24b}
\end{equation}
which is the diameter above which avalanches
become exponentially rare, and correspondingly
displacements by $\xi^\zeta$, durations by
$\xi^z$, etc., then {\it functions} such as
those that occur in the distribution of
avalanche sizes Eq~(\ref{eq:C18}), or the
average growth of the displacements during an
avalanche, $<\frac{\partial u(\vec
r,t)}{\partial t}|s>$ will be {\it universal}
functions of {\it scaled variables} such as 
$L/\xi$. For example, from Eq~(\ref{eq:D6}) we
obtain, for $K(\vec q, \omega)=-\eta i\omega
+D\eta|q|^\alpha$ and $d<d_c=2\alpha$,
\begin{equation}
<\frac{\partial u(\vec r, t)}{\partial
t}|s>\approx\frac{C_u}{C_t}\xi^{\zeta-z} Y\left(
\frac{\vec r}{\xi}, \frac{t}{C_t\xi^z}, \frac{s}
{C_u\xi^{d+\zeta}}\right)
\label{eq:D24c}
\end{equation}
with $s=\int d\vec r\Delta u (\vec r)$ the total
size, $C_u$ and $C_t$ {\it
non-universal} (dimensionfull) coefficients
which set the scales of the displacements and
times; these depend on the random pinning,
$\eta$,
$D$; etc. The {\it universal scaling function}
is 
\begin{equation}
Y(\vec R, T, m)=
\int_{\vec Q}\int_\Omega\int d\Lambda e^{i\vec
Q\cdot \vec R-i\Omega T}e^{-\frac{1}{2}m\Lambda
^2}\sqrt{\frac{m^3}{2\pi}}\left[ \frac
{\Lambda^2} {\Lambda^2 +(-i\Omega+|Q|^\alpha)^2}
\right]
\label{eq:D24d}
\end{equation}
which  depends {\it only} on the dimension, the
range of interactions, and the type of dynamics
(i.e. dissipative), as is manifested in the low
frequency form of the stress transfer function
$K(\vec q, \omega)$. As we shall see in the
next section, a similar scaling structure is
expected to exist in more realistic models.

Let us now try applying the toy model results to
the interface problem with $d=2$ and short range
elasticity, so that $\alpha=2$. This dimension is
less than 
\begin{equation}
 d_c^{short-range}=d_c(\alpha=2)=4,
\label{eq:D24e}
\end{equation}
 so we have $\zeta=2$,
i.e. $\Delta u(L)>>L$, for large avalanches. 
But this is clearly unphysical: our original
model for the interface assumed that it was
close to flat so that, at least on large scales,
we need small angles of the interface  i.e.
$\nabla u<<1$.  Thus the result Eq~(\ref{eq:D20})
violates the assumptions of our original model in
this case.

What has gone wrong? Is the original model bad or
have we made some grievous errors in trying to
analyze it?
The answer is the latter and understanding why
gives some clues as to how to do better.

In the infinite range mean field model, the odds
of any given segment jumping more than once are
very low as long as $s<<\sqrt{N}$, since the
odds of a specific segment jumping at all is
small and each jump is almost independent by
(justifiable) assumption.
But in the finite range models, we assumed that
this independence was still true, i.e. that the
probability of a segment $u(\vec r)$ jumping in a
short time interval depends only on the  {\it
increase} in pulling force, $\Delta\phi(\vec r)$,
on it during that interval.  But this is
problematic:  if a segment has just jumped, it is
much less likely to do so again until its
neighbors have caught up.  The needed increase in
$\phi(\vec r)$ for a subsequent jump will
typically be of order one. [Actually this is true
even in our toy model, but the cumulative effect
of many jumps causes the problem: the needed
$\Delta\phi(\vec r)$ to cause a large $\Delta
u(\vec r)$ that consists of many jumps should be
$\Delta\phi(\vec r)\approx\tilde {J}\Delta u
(r)\pm O(1)$ while in the toy model it is, for
$\rho=1$ where large avalanches can occur, 
$\Delta\phi(\vec r)\approx\tilde {J}\Delta u
(r)\pm
\tilde {J}\sqrt{\Delta u (r)}$.  This difference
is responsible for major errors when avalanches
involve large $\Delta u (r)$'s but {\it not} for
$d>d_c(\alpha)$ where $\Delta u (r)$'s remain of
order one.]

Our task, then, is to somehow take into account
properly the anticorrelations between local
susceptibilities to successive jumps. This will
certainly involve ensuring that   the
statistical properties of
$\{f_p[\vec r,u(\vec r)]\}$ when all $u(\vec r)$ 
are increased by any fixed amount are preserved;
i.e. the {\it statistical translational invariance}
of the system which is lacking in our toy
avalanche model.

Remarkably, in spite of its problems the toy
model {\it correctly} gives the statistics and
properties of avalanches for large $s$ and small
$\epsilon$ in dimensions greater than
$d_c(\alpha)$.  The basic reason for this is the
observation  mentioned above that each segment
is unlikely to jump many times during
even very large avalanches;  a real
understanding, however, relies on the renormalization
group treatment discussed briefly in the next
section.

\subsection*{VI. Interfaces and Scaling Laws}

Motivated by the partial success of the toy
model and general scaling concepts from more
conventional critical phenomena, we will now
approach the interface problem by making a {\it
scaling  Ansatz}. Specifically, we conjecture
that large avalanches near the critical point
have properties which scale with their diameter,
$L$, (or size $s$) as powers of $L$
\cite{Int1}\cite{Int2}. In the toy model we found that
the critical exponents which characterized these
scaling laws, $\zeta,\ d_f,\
\kappa \mbox{ and } z$ depend on the dimension
of the elastic manifold,  the power law decay
of the interactions if they are long range, and
on the type of dynamics, but {\it not on other
details of the system}.  This is the fundamental
property of ``{\it universality}''. In contrast,
the critical force and coefficients in the
 scaling functions such as Eq~(\ref{eq:D24c}),
will generally depend on details and hence are
non-universal.  We  might thus hope that in
dimensions
$d<d_c(\alpha)$ for which the toy model fails,
the exponents will still be universal functions
of $d$ and $\alpha$.

In addition to the four exponents already
introduced, there should also be an exponent
which characterizes the {\it correlation
length}: this is the diameter above which
avalanches become unlikely (like $\xi\sim
s_{max}^{\frac{1}{2\alpha}}
\sim\epsilon^{-1/\alpha}$  in the toy model)
\cite{MN}.  As
$F$ increases towards
$F_c$ with a generic past history, we conjecture
that
\begin{equation}
\xi\sim\frac{1}{(F_c-F)^\nu}.
\label{eq:E1}
\end{equation}
If in the mean field model the local jump
susceptibility $\rho$ is smooth at $F_c$
\cite{MidN} so
that 
$\epsilon\sim F_c-F$, then $\nu_{toy-
model}=\frac{1}{\alpha}$. This turns out to be the case for the
fuller mean field model defined in section III.  

It appears that we
have five separate exponents and we would like
to understand whether there are some relation
among them.  Fortunately, this will turn out
to be the case.
First, however, we can eliminate one exponent.

If, as we expect for $d<d_c$, $\zeta>0$ so that
some interface segments advance large distances in
big avalanches, then the elasticity will
certainly make neighboring regions advance
as well so that it seems implausible that
avalanches would be fractal; thus we expect
\begin{equation}
d_f=d
\label{eq:E2}
\end{equation}
for $d<d_c$.

\begin{figure}
\begin{center}
    \leavevmode
    \epsfxsize=3.5truein
    \epsfbox{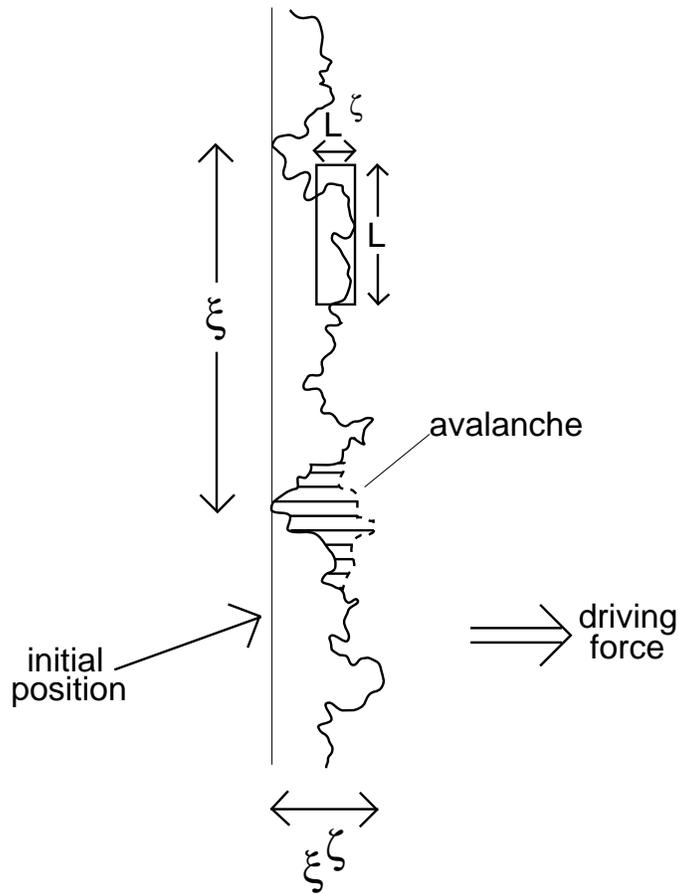}
  \end{center}
\caption{Schematic of a one-dimensional
interface with $F$ somewhat below the critical force.
After starting from a flat initial configuration, the
typical displacements of the interface are of order
$\xi^\zeta$ with $\xi$ the correlation length. On
smaller length scales, the statistics of relative
displacements as a function of separation has a simple scaling behavior
as 
shown. An avalanche
that occurs as the force is increased slightly is
also shown.}
\label{fig:5}
\end{figure}

The other relations between exponents are much
more subtle.  The simplest assumption---and one
that is borne out---is that there is one basic
length scale $\xi$ within the interface and a
basic scale 
\begin{equation}
\Delta u\sim\xi^\zeta
\label{eq:E3}
\end{equation}
in the direction of motion. [Note that we have
been sloppy and not put the needed non-universal
coefficients, as appeared in Eq~(\ref{eq:D24c})
into Eq~(\ref{eq:E3}).] Thus, for example, if we
started with a flat interface and gradually
increased
$F$, when
\begin{equation}
\epsilon\equiv F_c-F
\label{eq:E4}
\end{equation}
is small, the interface would be rough on scales
less than $\xi$, with
\begin{equation}
<[u(\vec r)-u(\vec r')]^2>\sim\left|\vec r-\vec
r'\right|^{2\zeta}
\label{eq:E5}
\end{equation}
for $\left|\vec r-\vec
r'\right|<<\xi$, and be flat on longer scales
with 
\begin{equation}
<[u(\vec r)-u(\vec r')]^2>\sim\xi^{2\zeta}
\label{eq:E6}
\end{equation}
for $\left|\vec r-\vec
r'\right|>>\xi$ as shown in Fig.~\ref{fig:5}. Similarly, we conjecture
that
for $\epsilon$ small, and $s$ large
\cite{Int2}\cite{BTW}
\begin{equation}
Prob(s)\sim\frac{1}{s^{B+1}}H\left(s/\xi^{d_f+\zeta}
\right)
\label{eq:E7}
\end{equation}
 with
$H(x\to 0)\to 1$ and
$H(x\to
\infty)\to 0$---i.e., a similar form to that  found in the toy
model. Scaling  implies that the exponent for
the distribution of avalanche probability as a
function of {\it diameter} in Eq~(\ref{eq:D22}) obeys
\begin{equation}
\kappa=B(d_f+\zeta)
\label{eq:E8}
\end{equation}
We can now compute the ``polarizability'' of the
system
\begin{equation}
\chi\equiv\frac{d<u(\vec r)>}{dF\uparrow}
\label{eq:E9}
\end{equation}
where the arrow denotes that $F$ is increasing.
This will be given by the sum over 
avalanches which are triggered with probability
$\rho dF$ per segment of the interface yielding
\begin{equation}
\chi \sim \rho\int\frac{ds}{s}\frac{s}{s^B} H(s/\xi^
{d_f+\zeta}).
\label{eq:E10}
\end{equation}
With $B<1$, this is dominated by large $s$ so
that
\begin{equation}
\chi\sim\xi^{(1-B)(d_f+\zeta)} \sim\epsilon^
{-(1-B)(d_f+\zeta)\nu}.
\label{eq:E11}
\end{equation}
But from the earlier results about the interface
roughness, Eq~(\ref{eq:E6}), we also have
\begin{equation}
\chi\sim\frac{d}{dF}\xi^\zeta\sim\epsilon^{-(1+\zeta\nu)}.
\label{eq:E12}
\end{equation}
We thus have 
\begin{equation}
B=\frac{d_f-\frac{1}{\nu}}
{d_f+\zeta} 
\label{eq:E13}
\end{equation}
and hence
\begin{equation}
\kappa=d_f-\frac{1}{\nu}.
\label{eq:E13a}
\end{equation}
\cite{Int2} Note that these relations work in the toy
model for $d>d_c(\alpha)$ but {\it not}  for
$d<d_c(\sigma)$ due to the problems discussed
earlier.

Another relation can be derived by considering
the forces one avalanche exerts on another
section of the interface.  When an avalanche that
moves a region of diameter $L\sim\xi$ occurs, we
expect it to also involve quite a few segments
nearby to jump, but not many that are far away. 
In particular, distances
$R\sim\xi$ away should be borderline.  The
increase in pulling force from the original part
of the avalanche on sections $\sim R$ away will
be
\begin{equation}
 \Delta\varphi\sim \xi^{d+\zeta}/R^{d+\alpha}
\label{eq:DE13b}
\end{equation}
(for short range elasticity, $\alpha=2$, the
argument is more subtle). For
$R\sim\xi$, this $\Delta\phi$ should be comparable
to the deviation $\epsilon$ from criticality or
else we must not have  properly identified the
crossover distance.  Thus we should have 
\begin{equation}
\Delta\varphi(R\sim\xi)\sim\frac{1}
{\xi^{\alpha-\zeta}}\sim\epsilon\sim\xi^{-\frac{1}
{\nu}}
\label{eq:E14}
\end{equation}
yielding the scaling law
\begin{equation}
\frac{1}{\nu}=\alpha-\zeta.
\label{eq:E15}
\end{equation}

A further relation can be derived by considering
the variations in the ``local critical forces''
$F_{cl}(\vec r, L)$ needed to make a region of
diameter $L$ advance by of order $L^\zeta$. We
define 
\begin{equation}
\epsilon_l(\vec r, L)\equiv F_{cl}(\vec r, L)-F
\label{eq:E16}
\end{equation}
to be, loosely, the deviation from "local
criticality".  Since the volume of the region
through which the  section of scale
$L$ will advance is $V\sim L^{d+\zeta}$, we
expect that the random pinning forces in the
region will make
\begin{equation}
\delta\epsilon_l(L)\equiv\sqrt{\mbox{variance}
[\epsilon_l (\vec r,
L)]}\geq\frac{1}{L^{(d+\zeta)/2}}
\label{eq:E17}
\end{equation}
i.e. the region ``does not know''  $F_c$
to better than this accuracy \cite{CCFS}.  If at scale
$L\sim\xi$, $\delta\epsilon_l (L=\xi)$ were much
larger than $\epsilon$, then we would expect many
regions of size $\xi$ to be unstable so large
avalanches that occurred in the past should have
been even larger. Thus we should have
\begin{equation}
\delta\epsilon_l(\xi)\leq\epsilon
\label{eq:E18}
\end{equation}
yielding
\begin{equation}
\frac{1}{\nu}\leq\frac{d+\zeta}{2} 
\label{eq:E19}
\end{equation}
from Eq~(\ref{eq:E17}).  This can be combined with
Eq~(\ref{eq:E15}) to yield
\begin{equation}
\zeta\geq\frac{2\alpha-d}{3}.
\label{eq:E20}
\end{equation}
An upper bound of  
\begin{equation}
\zeta\leq 2\alpha-d
\label{eq:E20a}
\end{equation}
  follows
from the observation that the toy model should
overestimate the jumps of a typical segment in a
large avalanche. For $d>d_c$, we expect $\zeta =0$;i.e. interfaces which are 
flat on large scales.

So far, we have not discussed the dynamics.  One
of the (helpful!) features of monotonic models,
is that at the end of an avalanche, how much
each segment has moved, $\Delta u(\vec r)$, is
{\it independent of the dynamics} although how
long the avalanche takes, in what order jumps
occur, etc. {\it will} depend on the dynamics
\cite{NoP}\cite{MidN}.  Indeed, even the exponent $z$
can depend on the stress transfer $J(\vec r,t)$.  In
particular for long range interactions whose effect is
only felt after a time proportional to $r$---i.e.
finite velocity of information propagation---one
must clearly have $z\leq 1$ (this can already
be seen in the toy model if $\alpha<1$) \cite{Ram3}. 
In general, however, we cannot say much about $z$
without much more work.

Nevertheless, using our scaling Ansatz, we can
relate the dynamical behavior in the moving
phase for $F$ just above $F_c$ to that for
$F<F_c$ \cite{Int1}\cite{Int2}\cite{CDWF}.  In
particular, we conjecture that the jerkiness of the
motion occurs on length scales up to
$\xi\sim\frac{1}{(F-F_c)^\nu}$ and times up to 
\begin{equation}
\tau_\xi\sim\xi^z
\label{eq:E20b}
\end{equation}
 while the motion is
smoother on longer scales.  This jerky motion
will look like the dynamics within avalanches. If
we consider, crudely, the motion to be made up of
avalanches of scale
$\xi$ occurring at intervals of order
$\tau_\xi$   within each region of
diameter $\xi$, then the velocity is simply
\begin{equation}
\overline{v}\sim\frac{\xi^\zeta}{\tau_\xi}\sim
(F-F_c)^\beta
\label{eq:E21}
\end{equation}
with \begin{equation}
\beta=(z-\zeta)\nu.
\label{eq:E22}
\end{equation}
Note again, that with the exponents of the toy
model  for $d>d_c$: $z=\alpha$,
$\nu=\frac{1}{\alpha}$, $\zeta=0$ and $\beta=1$,
the scaling law Eq~(\ref{eq:E22}) is obeyed.

In the moving phase, the response  of $<u(\vec
r, t)-\overline{v}t>$ to a small additional {\it
static} force $\delta F_qe^{i\vec q\cdot \vec
r}$,
\begin{equation}
\chi(\vec q)=\frac{\delta<u(\vec q,
\omega=0)>}{\delta F_q},
\label{eq:E22a}
\end{equation}
can be obtained exactly! [Note that here either
sign of $\delta F_q$ is okay if it is
sufficiently small.] If the variable  change
\begin{equation}
u(\vec r,t)=\tilde{u}(\vec r, t)+ \int_{\vec q}
e^{i\vec q\cdot\vec r}\frac {\delta F_q}
{\tilde{J}-J(\vec q,\omega=0)}
\label{eq:E22b}
\end{equation}
which corresponds to transforming away the
response in the absence of pinning, is made,
then the {\it statistical properties} of the
random pinning forces 
\begin{equation}
\tilde{f}_p[\vec r,
\tilde{u}(\vec r, t)]\equiv f_p[\vec r,u(\vec r,
t)]
\label{eq:E22c}
\end{equation}
as a function of $\tilde{u}$ and $\vec r$ are
{\it identical} to those of the original  
$f_p$. This is a consequence of the
 underlying statistical rotational invariance of
the system \cite{Aniso}.  We thus find that 
\begin{equation}
<\tilde{u}(\vec r,t)>_{\delta F_q} =
<u(\vec r, t)>_{\delta F_q=0}
\label{eq:E22d}
\end{equation}
so that the average response is given exactly by
the second term in Eq~(\ref{eq:E22c}), i.e.
\begin{equation}
\chi(\vec q)=\frac{1}{\tilde{J}-J(\vec
q,\omega=0)}\sim\frac{1}{|q|^\alpha}.
\label{eq:E22E}
\end{equation}
Since $\chi$ should scale as
$L^\zeta/L^{\frac{1}{\nu}}$, we again obtain the
scaling law Eq~(\ref{eq:E15}). Note that since
$\zeta$ is defined at $F_c$, the agreement of
this result with that below threshold supports
the notion that the appropriate characteristic
length scales above and below threshold will
diverge at $F_c$ with the {\it same} exponent
$\nu$. 

We have reduced the number of critical
exponents to two basic ones; $\zeta$ and $z$
which relate displacement and time scales to
length scales. But, so far, we have neither a means
of computing these exponents; nor more
importantly, a way of justifying the scaling laws
beyond the hand waving arguments given above (or
variants of these); nor even a way of
understanding the claimed universality of the
exponents.

But with the   basic scaling picture in mind, we
can try and appeal to the framework of the
renormalization group which has been so
successful in understanding equilibrium---and
some non-equilibrium---critical phenomena. There
are two substantial difficulties. One is
associated with the basic physics: we must find
a way of properly dealing with two kinds of
scales in the pinned phase near $F_c$.  First,
the jumps of small segments happen on the basic
microscopic time scale but their existence and
discrete nature is crucial. Large
avalanches last for times which scale with their
size out to $\tau_\xi$ which is very large just
below $F_c$ and thus avalanche activity spans a broad range of scales.  But, in 
addition, there is the time {\it between}
avalanches set by the rate at which $F$ is
changed; as long as this is slow enough, it does
not really matter except that the important
anti-correlations between successive avalanches
in the same region will only be felt on this
very long time scale.

The other main difficulty is associated with the
history dependence in the pinned phase.  For
example, what should one average over to get
sensible quantities? If the stress transfer, $J
(\vec r,t)$, is non-negative, this difficulty can
be circumvented if $F$ is always
increasing (or always decreasing): if this is the
case, then from any stable initial condition the
pulling force  on every segment will increase
monotonically with time.  This feature is important for the physics as well as 
drastically
simplifying the theoretical analysis.
Nevertheless, even in such monotonic models, there will be history dependence. 
But near to $F_c$, this should, at worst, only modify non-universal coefficients 
as long as the critical force is approached monotonically starting from a much 
lower force. A natural reproducible history results from starting at $F=-F_c$.

For monotonic models, a perturbative
renormalization group $(RG)$ analysis for
dimensions near the upper critical dimension has
been carried out \cite{Int1}\cite{Int2}\cite{EK}. The
first result  is that for
$d>d_c(\alpha)$, the decreased local
susceptibiltiy to jumping after a segment has
jumped is {\it irrelevant} in the $RG$ sense,
except on the very long time scales during which
the whole system has advanced.  In particular,
this justifies our claim that, while the
critical force and other non-universal
properties will not be given correctly, {\it
universal} features of avalanches such as
critical exponents and scaling functions [e.g.
$H$ in Eq~(\ref{eq:E7}) and $Y$ in
Eq~(\ref{eq:D24c})] {\it will be given exactly} by
the toy model for
$d>d_c$.  In the moving phase, some of the effects left
out of the toy model will be important but mean
field results like $\beta=1$  and $\zeta=0$ will obtain
\cite{MFB}.  

For
$d<d_c$ many new results can be obtained from the RG analysis.  In addition
to the derivation of universality, scaling laws
and perturbative computations of exponents that
arise from a new {\it critical fixed point} of the
$RG$, it is possible, in principle---although not
yet carried out---to compute such quantities as
the new universal scaling function that replaces
$Y$ in Eq~(\ref{eq:D24c}), anticorrelations
between avalanches in the same region, local
velocity correlations in the moving phase, etc.

Here we just quote the results for the
exponents \cite{Int1}\cite{Int2}\cite{EK}.  All the
scaling laws derived heuristically above are found to
be obeyed. The exponent $\zeta$ is
\begin{equation}
\zeta\approx\frac{d_c(\alpha)-d}{3}
\label{eq:E23}
\end{equation}
to all orders in powers of $d_c-d$; indeed, this
result---which saturates the lower bound
Eq~(\ref{eq:E20})---may well be exact. Numerical
computations for $\underline{d=\alpha=1}$
yield 
\begin{equation}
\zeta\approx 0.34\pm0.02
\label{eq:E25}
\end{equation}
consistent with $\frac{1}{3}$
\cite{Ram3}\cite{SMN}.  The dynamic exponent
is found to be
\begin{equation}
z=\alpha-\frac{2[d_c(\alpha)-d]}{9}+O[(d_c-d)^2].
\label{eq:E26}
\end{equation}
We thus find an interesting effect: the
non-linearities of the avalanche process cause
disturbances of the interface to propagate more
{\it rapidly} at long scales than for an
unpinned interface.  The velocity exponent is
then, from scaling,
\begin{equation}
\beta=\frac{z-\zeta}{\alpha-\zeta}\approx1-
\frac{2(2\alpha-d)}{9\alpha}<1
\label{eq:E27}
\end{equation}
i.e. a concave {\it downwards} $\overline{v}(F)$
curve.  Making the somewhat dangerous
extrapolation to the {\it interface} with
$\underline{d=\alpha=2}$,
we get predictions of
\begin{equation}
z\approx\frac{14}{9},\ \zeta\approx\frac{2}{3}
\mbox{,  and  }\ \beta\approx\frac{2}{3}.
\label{eq:E27b}
\end{equation}

At this point, except for some experiments on
interfaces between two fluids that are being
driven through porous media which give somewhat
inconclusive results \cite{PorE}, there are not
experiments, of which this author is aware, that test
these results for interfaces.  But as we shall see in
the next section, there have been both experiments
and numerical tests   carried out on other
systems.

\subsection*{VII. Applications and Complications}

In the previous section we saw how scaling ideas
and intuition that arise  from the simple solvable
toy model could be used to develop an
understanding of the behavior of interfaces near
to the critical driving force that makes them
move.  Renormalization group methods can then be
used to carry out concrete calculations and
justify many of the conjectures.  In particular,
the general {\it structure} and existence of
scaling laws and universality follows rather
directly from the existence of an $RG$ critical
fixed point.

One of the advantages of this framework is that
it enables us to apply the general
structure to other systems---such as (but not
limited to) different dimensionalities and
ranges of interactions. But in addition we can introduce
various physical features left out of the simple
models---even our relatively realistic
Eq~(\ref{eq:B2})---and ask whether they are
{\it relevant} in the $RG$ sense of changing (or
destroying) the universal aspects of the critical
behavior. In this section we will discuss several
of the physical systems mentioned in the
Introduction with an eye both to applying the
ideas and seeing how they must be modified---or
thrown away!---to account for the
appropriate extra physics.

\subsubsection*{A. Charge density waves}
\begin{figure}
  \begin{center}
    \leavevmode
    \epsfxsize=3.5truein
    \epsfbox{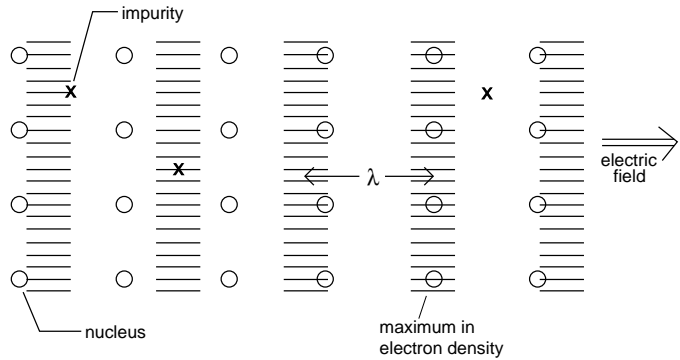}
  \end{center}
\caption{Schematic of a crystal with a
charge density wave is incommensurate with
respect  to the underlying lattice periodicity of the
solid. The maxima in the difference
between the actual electron charge density and that without the
CDW are shown. An electric field pulls on the CDW
which could move freely,except for being impeded by pinning forces of
the impurities.}
  \label{fig:6}
\end{figure}

The best and perhaps really the only real test
of  critical behavior near to a depinning
transition of an elastic manifold in a random
medium is that of charge density waves (CDW)
driven by an electric field \cite{CDW1}.  The periodic
electron density waves that occur within this
class of materials are incommensurate with respect
to the underlying crystalline periodicity in one
direction so they could move freely in this
direction---contributing to the current
proportionally to their velocity---except for
being pinned by randomly positioned impurities;
see Fig~\ref{fig:6}.  These CDWs are thus three dimensional
elastic manifolds with short range interactions
(the coulomb interactions are screened) in a three
dimensional random medium.  Inertial effects are
negligible.  The primary difference from
interfaces is in terms of the displacements
$u(\vec r)$: in the frame of the crystal the
random pinning forces $f_p[\vec r, u(\vec r)]$ are
{\it periodic} in 
\begin{equation}
u(\vec r)\to u(\vec
r )+\lambda_{CDW} 
\label{eq:Fa1}
\end{equation}
 for all $\vec r$, where
$\lambda_{CDW}$ is the wavelength of the CDW. 
Although the resulting behavior near the
critical driving force is not much changed for
$d>d_c=4$, the extra symmetry associated with
this periodicity changes the {\it universality
class} for $d<d_c$.  In particular, the
exponents become \cite{CDWNF}
\begin{eqnarray}
\zeta=0, \nonumber\\
\nu=\frac{1}{2},\ \mbox{ and }\nonumber\\
z=2-\frac{4-d}{3}+O(4-d)^2,
\label{eq:Fb1}
\end{eqnarray} 
 yielding scaling behavior
above $F_c$ involving the velocity exponent
\begin{equation}
\beta=z\nu=1-\frac{1}{6}(4-d)+O(4-d)^2\approx\frac
{5}{6}
\label{eq:F1}
\end{equation}
in $d=3$. Experiments \cite{Bhat} carried out on CDWs 
and numerical simulations yield
$\beta\approx 0.75-0.9$ over $2\frac{1}{2}$
decades of
$F-F_c$---surprisingly good agreement with the
theoretical prediction.

But should we really expect that CDWs will
exhibit the critical behavior of an ideal
elastic medium? In a precise sense, certainly
not. {\it Thermal fluctuations} (or even
quantum fluctuations) can cause sections of the
CDW to overcome the barriers caused by the pinning and
jump to lower energy local minima.  For any
non-zero electric field, this will cause the CDW
to gradually creep forwards thereby contributing
to the current.  The critical behavior near the
fluctuationless $F_c$ will thus be smeared out by
fluctuations which are hence a {\it relevant}
perturbation \cite{CDWF,MidTh}.  This is quite analogous to
the role of a magnetic field near to ferromagnetic
phase transitions, which also smears out the
critical singularities. But as in the magnetic
case, if the perturbation is small enough,
critical behavior can still be observed over a
wide range of scales and $F-F_c$ with the 
smearing only occurring very close to $F_c$. In
general,  fluctuation effects appear to be
quite substantial in CDW systems and the
critical behavior is smeared out.  But the
experiments quoted above \cite{Bhat} were performed by
applying an ac driving force in addition to the
dc drive; this reduces the effects of thermal
fluctuations and appear to yield a wide range over which
they do not play much of a role.

Another complication in CDWs---which might
also be reduced by an additional ac drive---is
{\it defects} in the CDW lattice, especially
dislocations \cite{Def}.  The pinning is very weak in
CDWs implying that the stresses that would cause
dislocations to form are unlikely to occur except
perhaps on long length scales.  In this weak
pinning regime, the effects of dislocations are
poorly understood. It now appears, however, that
dislocations will {\it not} destroy the existence of the
CDW phase in {\it equilibrium} \cite{BrG}; this is not
directly relevant to the non-equilibrium physics of
interest here but may point to progress also on
the more relevant issues.  Nevertheless, at this point,
whether or not defects always destroy the elastic
depinning critical behavior that we have studied
is an open question.

\subsubsection*{B. Superconductors}
One system that is quite similar in spirit to
a CDW, but for which the strength of the pinning
forces can readily be varied , is a {\it vortex lattice}
in a type II superconductor.  Vortex lattices are
pinned by impurities that impede their motion
under the action of a transport current which
exerts a force on the vortices \cite{Blat}.  Since the
voltage is proportional  to the mean vortex
velocity, $\overline v(F)$ plots
are simply voltage-current curves, $F_c$
being simply proportional to the {\it critical current}
density.

By making bulk superconducting samples, thin
superconducting films or wires, or a normal
layer sandwiched between two bulk
superconductors, three, two and one 
dimensional systems can all be studied as well
as, in the last case, a two dimensional lattice
of roughly parallel vortex lines  with
no dislocations allowed.  In principle, many
systems and regimes can thus be investigated,
subject to the complication of non-uniform forces
on the vortices, dissipative heating effects, and
 natural tendencies of experimentalists to care
more about the magnitude of the critical current
density  than about what happens when the
superconductor  fails, i.e. when the critical
current is exceeded!  Surprisingly, although
thermal fluctuation driven transitions in both
clean and dirty superconductors have received a
lot of attention recently, \cite{Blat}\cite{FFH} the
behavior nearer to critical currents has received far
less \cite{CrC}.

\begin{figure}
  \begin{center}
    \leavevmode
    \epsfxsize=3.5truein
    \epsfbox{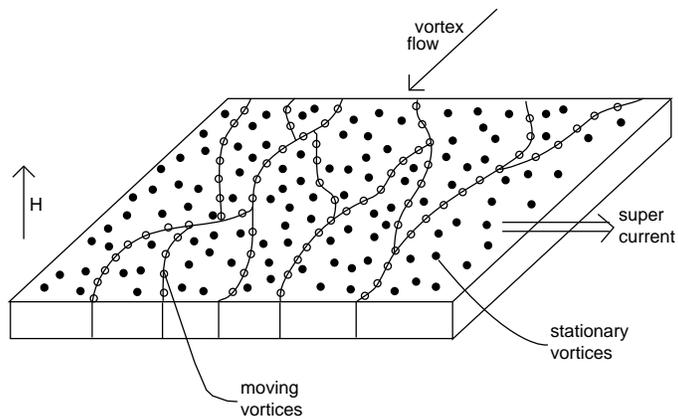}
  \end{center}
\caption{A thin superconducting film in a
magnetic field , $H$, perpendicular to
the film. A supercurrent provides a driving force for
the vortices to move accross  the film. Just above
the critical current for vortex motion, the vortices 
in most of the film can remain stationary with the
vortex flow confined to a sparse network of vortex
channels, as shown. The electric field is
proportional to the average vortex flow rate.}
  \label{fig:7}
\end{figure}

If the pinning is very strong, the vortex
lattice will be destroyed.  What then (when
thermal fluctuations can be neglected) will be
the qualitative behavior near the critical
force? In the case of two dimensional films in a
perpendicular magnetic field that produces point-like
vortices,  the vortex flow  just above $F_c$ will
almost certainly  be confined to a {\it sparse}
interconnected {\it network} of irregular  {\it
channels} across the system, as sketched in
Fig~\ref{fig:7}.  Some preliminary theoretical studies \cite{Chan} and
experiments of
the critical behavior that can arise in this
regime have been carried out  and experiments that "see" the vortices performed 
\cite{VorIm}, but even this
simple case is far from understood. For more
complicated cases of intermediate strength
pinning or when three dimensional effects are important,
even less is known.

  One thing that is clear, both theoretically
\cite{Chan} and experimentally \cite{CrC}, is that
the critical force is history dependent.  This and
other history dependence---which can be caused by
defect motion---is certain to play an important role in
the physics.  We note that in principle---and
soon, if not quite now, in practice \cite{VorIm}---measurements
of local vortex flow on small scales will be
possible in some of the parameter regimes of
greatest potential interest. Both macroscopic and
microscopic information should be available in
these superconducting vortex systems.

Leaving the effects of lattice defects as an
intriguing puzzle, we now turn to another type
of physical effect that we have so far left out:
the role of inertia and other related phenomena.

\subsubsection*{C. Cracks}
\begin{figure}
  \begin{center}
    \leavevmode
    \epsfxsize=3.5truein
    \epsfbox{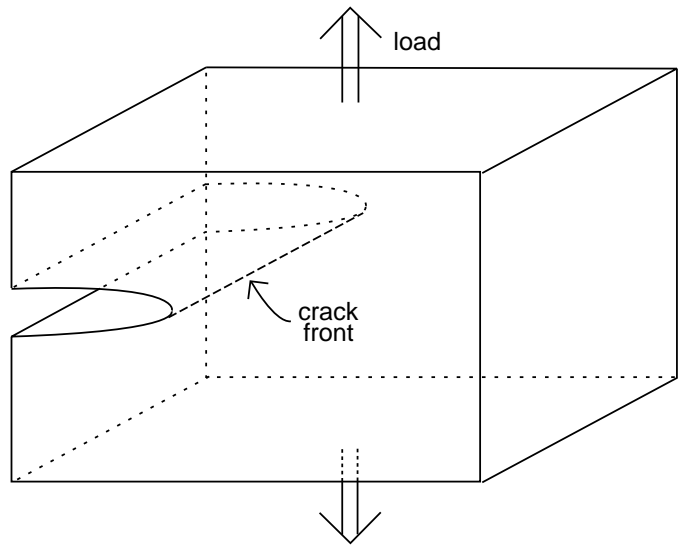}
  \end{center}
\caption{ A solid with a crack under
tensile loading. If the load is increased, the crack
front can progress through the solid.  In some circumstances, the crack is 
confined to a plane; this is the case discussed in the text.}
  \label{fig:8}
\end{figure}

In contrast to CDWs and vortex lattices,  interfaces between phases are
``topological'' and hence have elasticity which is
robust for weak pinning, although for strong
pinning they can become fractal and a very
different ``invasion percolation'' behavior 
occurs. Various other elastic systems are
also {\it not} directly susceptible to destruction
by defects.  In particular, the front of a {\it planar
tensile crack} in a heterogeneous solid has long
range
$\frac{1}{r^2}$ elastic forces mediated by the
elasticity of the solid which act to keep the
crackfront roughly straight; see Fig~\ref{fig:8}
\cite{Ram1,Ball,Fre}.  But random variations in the  local
toughness---the energy per unit area needed for crack
growth---act to deform the crack front.  This
system is thus an example of an elastic manifold with
$d=1$ and $\alpha=1$. 

Numerical studies \cite{Ram3}\cite{SMN} with
quasistatic (i.e. instantaneous on the time  scales of
the crack motion) stress transfer which corresponds to
$J(r,t)\alpha\delta(t)/r^2$ and  hence is a
monotonic system, yields results for $\zeta,\ z,\
,\nu,\ \mbox{and }\beta$ in excellent agreement
with the $d=2-\epsilon$ expansion results \cite{EK}. 
But the
resulting $z\approx\frac{7}{9}$ is less than one \cite{Ram3,EK}.
Therefore, even if the basic propagation of
disturbances along the crack front is slow on
small scales  (so that the quasistatic
approximations would seem to be justified), near
the critical point disturbances would propagate
arbitrarily fast since the characteristic scale of
the ``velocity'' along the crack front would
scale  as
$\xi/\xi^z$ which diverges as
$\xi\to\infty$. This is clearly unphysical and so
elastic wave propagation effects {\it must}
change the asymptotic critical behavior.  The simplest way
to take the time delays associated with wave
propagation into account is to use
\begin{equation}
J(r,t)\sim\frac{1}{r^2}\delta(t-r/c)
\label{eq:F2}
\end{equation}
with $c$ an elastic wave velocity. This is still a
monotonic model, so only the dynamic
exponent $z$ will change. Theoretical arguments
and numerical simulations \cite{Ram3}
 for this case yield $z=1$ exactly---so there is
no problem with causality. From
Eq~(\ref{eq:E22}), we obtain
$\beta=1$.

But real elastic wave propagation is much more
complicated \cite{Ram3}\cite{Fre}. If  the crack
grows by a segment of
the crack front jumping forward only to be stopped by a tougher region,
then a point on the crackfront a distance $r$
away will initially feel no change in stress. Furthemore, when the
longitudinal waves with velocity $c_l$ arrive,
the stress that tends to open the
crack---analogous  to
$\sigma(r,t)$ for the interface model---will
actually {\it decrease}   since
$J(r,\ t \raisebox{-.6ex}{
$\stackrel{\textstyle>}{\sim}$ } r/c_l)<0$
\cite{Ram2,Ram3}\cite{MouW}! Only when the Rayleigh waves
with slower velocity
$c_R\leq c_t< c_l$ arrive will the opening
stress become positive. But when this occurs,
there will  be a rapid jump to a {\it large peak}
value of the opening stress followed by a gradual
fall  off   to the eventual static stress
increase which is   
\begin{equation}
J(r, \omega=0)=\int^\infty_0
J(r,t)dt\sim\frac{1}{r^2}.
\label{eq:F3}
\end{equation}
This is highly non-monotonic behavior! Such
{\it stress overshoots} can cause segments of the crack
front to jump forward that would {\it not} have
been triggered by the static stress changes. How
often this occurs will depend on the heights of
the stress peaks relative to the static stress
increases---i.e. to the size of the overshoots.
If the jumps of segments of the crack are slow and
smooth (with duration long compared to the sound
travel time across the jumping segment) then the
stress overshoots will only occur far from the
jumping segment and have small amplitude. Their
effects should then be small and the quasistatic
behavior should be observable---except, as we
shall see, very near
$F_c$. But {\it any} stress overshoots will at
least occasionally cause {\it some} extra jumping
and we must understand their effect. 

Some intuition can be gleaned by considering what
would happen if the stress increases {\it never}
decayed to their static value, but stayed at
their peak strength. This would correspond,
roughly, to increasing the elastic interactions
of the crack with  itself. As we saw in the mean
field model, such enhanced elasticity causes the critical
force to decrease because of more effective averaging over the randomness [see 
Eq~(\ref{eq:B16})]. Thus
we would expect avalanches to run away at a
force {\it below} the quasistatic $F_c$.
A careful analysis of the effects of realistic
stress transfer peaks followed by decay towards
the static stress shows that the same basic
picture obtains \cite{EQ2}\cite{Ram3}. 

Consider a
segment which, at the end of the quasistatic
approximation to an avalanche, has a final stress on
it which is very close to being enough to make it jump
again. Then any overshoot in the stress on this
segment that occured during the actual avalanche
will, if the resulting peak stress is above the
quasistatic final stress, cause it to undergo an
extra jump. If, as is indeed the case, there is a
small density of these extra jumps roughly
randomly distributed within the avalanche, each
of the extra jumps will trigger extra avalanches,
for which one again has to consider the effects of
the stress overshoots which  make more
avalanches, etc. One can show that this process 
will {\it always} run away for a sufficiently
large avalanche---independent of how small the
stress overshoots might be \cite{Ram3,EQ2}.

The stress peaks are thus a {\it relevant}
perturbation which will change the critical
behavior.  Indeed, the runaway of large avalanches implies
the  {\it coexistence of moving and stationary
states} and strongly suggests that the depinning
transition in the presence of stress pulses will
be either {\it first order} with a discontinuous
and probably hysteretic  $\overline{v}(F)$ or have a
continuous $\overline{v}(F)$ but with a smaller
critical force and different critical behavior
\cite{Ram3}.

What happens if the stress overshoots are
small, say with strength parameterized  by
$\Sigma$? Then only avalanches with
\begin{equation}
L\geq
L_\Sigma\sim\frac{1}{\Sigma^{1/y_\Sigma}}
\label{eq:F4}
\end{equation}
will typically be much affected by the stress
peaks. The exponent $y_\Sigma$  is the RG
eigenvalue for the relevant perturbation,
$\Sigma$; it has been computed analytically \cite{EQ2,Ram3} and
checked numerically \cite{Ram3} for some types of stress
overshoots, but it's value is not yet understood
analytically for the type of long-tail stress
peaks that occur in the crack problem
\cite{Ram3,MouW}. 

Experiments on very slowly advancing cracks
confined to a plane \cite{SME} yield an estimate of
the roughness exponent which characterizes the
deviations  of the crack front  from a straight
line, of
\begin{equation}
\zeta_{crack\ front}\approx 0.55\pm.05
\label{eq:F5}
\end{equation}
substantially larger then the quasistatic
prediction of $\frac{1}{3}$. However the large 
corrections to scaling predicted by the RG
analysis \cite{Ram3}\cite{NPriv} may account for this
discrepancy; beware of quoted error bars for
exponents! 
It is possible, however, that real
elastodynamic effects play a role even for
slowly moving planar cracks.

A related challenge is to understand  the crack front
distortions which manifest themselves in the
roughness of the {\it fracture surface} left behind
after a crack front passes: i.e. after the
material is broken \cite{Bou}\cite{Ball}\cite{Ram2}.
Again, elastodynamic effects may play a crucial role. In certain situations, 
cracks can form multiple branches  --such as shattering of glass--, or just a 
few small branches near the crack front \cite{Fre}. Under what circumstances 
crack branching effect might affect the onset of crack motion or the large scale 
shape of fracture surfaces is another open question.

\subsubsection*{D. Faults}

So far---except for 
crack front roughness---we have only discussed
systems for which the simple  measurements are of
macroscopic properties such as
$\overline{v}(F)$. But there is one natural
system in which, even though the length scales
are huge, the measurements are of
``microscopic'' type: specifically, the
statistics and other properties of
``avalanches''.

\begin{figure}
  \begin{center}
    \leavevmode
    \epsfxsize=3.5truein
    \epsfbox{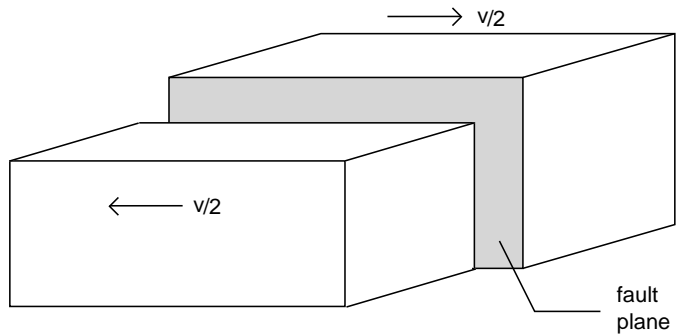}
  \end{center}
\caption{Schematic of two segments of
crust which move relative to each other along a fault
plane in a sequence of earthquakes. The driving forces
are transmitted to the fault by the two halves of the
crust moving with relative velocity, $v$.}
  \label{fig:9}
\end{figure}

A crude model of a geological fault is the motion
of two large blocks of crust in contact along a
disordered but roughly planar surface---the
fault---which move relative to each other; see
Fig~\ref{fig:9}.  The motion is driven by forces exerted
from far away (e.g. from the viscoelastic region
beneath the crust) that are transmitted to the
fault by the elasticity of the blocks. Rather
than being driven by a constant force, the blocks
are driven at a fixed time-averaged velocity that
is extremely slow---of order millimeters to
centimeters per year---compared to other
characteristic velocities---e.g. the speed of
sound in the rock \cite{EQ1}.  Segments of the two
dimensional fault plane interact with each other
by long range
$\frac{1}{r^3}$ stress transfer --as can be seen by dimensional analysis-- and 
are pinned
by heterogeneities on the two fault surfaces
that have to rub past each other \cite{EQ2}. 

Earthquakes,
of course, are just the ``avalanches'' that occur
when some segment of the fault becomes unpinned
and jumps forward triggering others
\cite{FFF,BEQ}. The range of length scales of
slipping regions is huge: from meters to hundreds of
kilometers, with slips---changes in relative
displacement across the fault plane, i.e.
$\Delta u$---from millimeters to tens of
meters. 

Most theoretical approaches to modelling
faults have involved physics driven by friction
laws and inertial effects \cite{CLS}, with intrinsic
heterogeneities playing a secondary role if
included at all. An alternative approach \cite{EQ2},
motivated by the systems we have discussed here,
might be to start from the opposite end, a
strongly disordered fault with quasistatic stress
transfer, and then bring in other features such
as elastic wave propagation (which is the
manifestation of inertia) and frictional
weakening. This has the advantage of building on
an established theoretical framework in which the
importance---or lack thereof---of various
features can be considered. This is particularly
useful as the parameter space in even
relatively simple models is very large and thus
hard to explore numerically---at least in ways
that might convince a skeptic of the
interpretations or predictions!

If the forces that one side of the fault exerts
on the other are independent of the history and
the local slip velocities
$\frac{\partial u(\vec r,t)}{\partial t}$  and
the stress transfer is  quasistatic, i.e.
  $J_{qs}\sim\delta(t)/r^3$,  then
this system
 falls into the class of
generalized interface models 
with $d=2$ and $\alpha=1$. But for this
$\alpha$, two is exactly the upper critical
dimension so we can use the results of the
toy model. The fault system acts as if it were
driven just below
$F_c$ (by an amount that goes to zero for large
system size) so that we can set $\epsilon=0$ \cite{SOC}.  [Note that there 
should be
logarithmic corrections to various quantities, as
is usual at critical dimensions, but their
effects are small].  The toy model is
sufficiently simple, that one can include a
realistic way of driving the faults, but for now
we consider the simpler idealized infinite system.

Several results can immediately be used \cite{EQ2}:
the area of the  region of the fault that slips in an
earthquake will scale as the square of its
diameter, $L$, the typical slip in this region
will depend only logarithmically on $L$ (i.e.
$\zeta=0$), and the size, in this context called the {\it moment}, will scale as
\begin{equation}
M=\int d\vec r\Delta u (\vec r)\sim L^2.
\label{eq:F6}
\end{equation}
 Note that
the {\it moment magnitude} often quoted in
newspapers is
\begin{equation}
m=\log_{30}M+\mbox{constant},
\label{eq:F7}
\end{equation}
the base 30 being for historical reasons,
particularly consistency with the older Richter
magnitude scale.
The probability of different sized events in the simple model is
\begin{equation}
{\rm Prob}\ (moment > M)\sim\frac{1}{M^B}
\label{eq:F8}
\end{equation}
 with $B=\frac{1}{2}$.

How do these predictions compare with
observations? The moment of quakes  is the best
measured quantity, although for medium  and
larger size events, the duration $\tau$ can also
be measured well.  The diameter---or generally the
dimensions of the slipped region---and the slip
$\Delta u$ can only be measured directly if the
quake involves slip  where the fault breaks
through the surface of the earth.
However, earthquake data are usually interpreted
in terms of a {\it crack picture} in which
$\zeta=1$ and $z=1$ so that
\begin{equation}
M_{crack}\sim L^3\sim\tau^3
\label{eq:F9}
\end{equation}
\cite{EQ1}. This scaling appears to be reasonably well
justified observationally for large earthquakes and the $M\sim \tau^3$
scaling perhaps also for intermediate size events. But
for small events---i.e. most of them---only the
moment can be measured reasonably reliably.
Thus, perhaps, our $M\sim L^2\sim \tau^2$ might
not be ruled out for small events although it probably
is for large ones. Note however, that in quakes
of magnitude $m=7$ or so and above, the linear size
(e.g. the diameter) is comparable to the depth
of the crust (and other relevant length scales) so
different scaling
laws should in any case be expected for large
earthquakes with a crossover between the two
regimes \cite{Pack}. 

\begin{figure}
  \begin{center}
    \leavevmode
    \epsfxsize=3.5truein
    \epsfbox{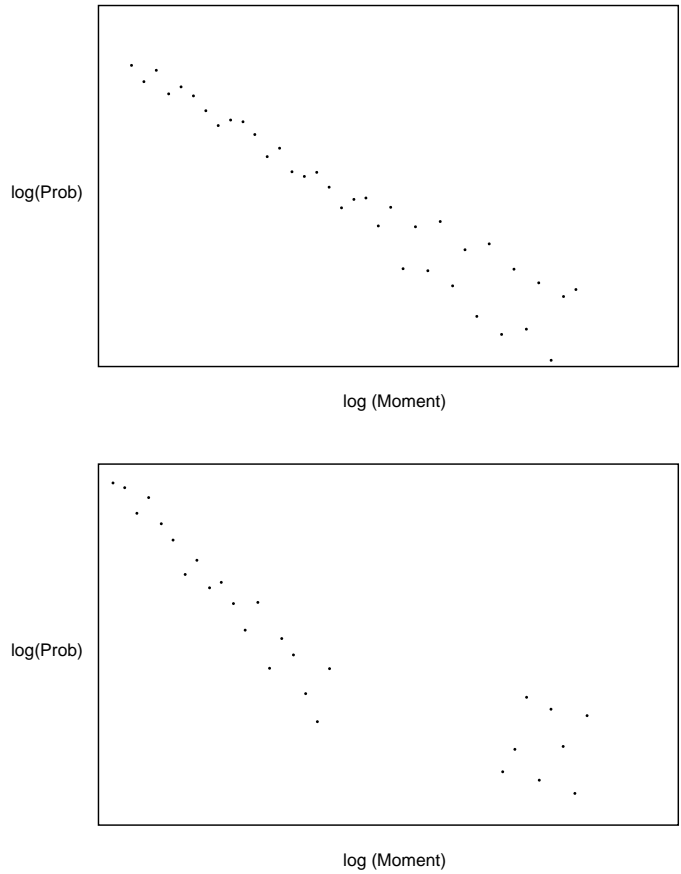}
  \end{center}
\caption{ Schematic of two types of earthquake
statistics observed on different geological faults; see [20] .
The probability of quakes with a given moment is
plotted on a log-log scale. a) A fault with power law statistics of events.
b) A fault that exhibits ``characteristic
earthquake'' behavior which refers to the peak in the
distribution for large events  of a characteristic
size, with not many intermediate size events occuring. }
  \label{fig:10ab}
\end{figure}

The question of earthquake magnitude
statistics is a complicated and controversial
one. The famous Gutenberg-Richter law \cite{GR}
states that
 the distribution of {\it all} earthquakes
approximately satisfies Eq~(\ref{eq:F8}) with a
$B\approx\frac{2}{3}$, although it changes
somewhat---perhaps for the reasons mentioned
above---around magnitude seven. The data cover
over twelve orders of magnitude in the moment,
equivalent to, assuming the crack scaling
Eq~(\ref{eq:F9}), four orders of magnitude in
length scale.
Understanding the Gutenberg-Richter law is a
very interesting problem, tied closely to that
of understanding the apparent fractal
distribution of some geological features such as
fault networks \cite{BEQ}.  But our
subject here is individual faults---at least if they
really exist as well defined entities, a question which
is also controversial. Observations of some
highly disordered faults have been fitted---over
a reasonable range---with $B\sim 0.5-0.6$
\cite{Wes}. This is perhaps encouraging although the
apparent rough agreement with our Eq~(\ref{eq:F8}) may
well be fortuituous. Most faults, however, exhibit
quite different behavior: a small regime (if at all)
of power law statistics that can be fit by
somewhat larger $B$'s, then a gap with few
events and a narrow peak (which dominates the
total slip) at a {\it characteristic earthquake}
size in which the whole fault section slips
\cite{Wes}.  This behavior, sketched in Fig~\ref{fig:10ab}b,
appears to be very different from the pwer-law scaling behavior
we have found in our simple quasistatic heterogeneous
model.  However, behavior  qualitatively similar
to Fig~\ref{fig:10ab}b---with large characteristic
earthquakes---has been found in simulations of
models with inertia and frictional weakening but
no intrinsic randomness although analytic
understanding of these results is rather limited
\cite{CLS}.

 Can we understand qualitatively how both
power-law and characteristic-earthquake types of
behavior might arise starting from our simple
randomness-dominated picture?
Elastodynamic effects, as for the crack front
problem discussed earlier, will result in peaks
in the dynamic stress transfer that are larger
than the static stress transfer. In addition,
frictional weakening---the tendency for dynamic
frictional forces to be less than static
frictional forces---will also be present. Both
of these effects will cause extra segments to
slip that would not have slipped in quasistatic
events. These, as for the crack front, will cause
runaway of large earthquakes which will eventually be stopped only
by strong ``pinning'' caused by  the
boundaries of the fault section, or by the
unloading of the shear stress that is driving the
fault. If the stress overshoot and frictional
weakening effects are small, there should be a
wide regime of power law scaling with
$B=\frac{1}{2}$ which can extend out to the
largest quakes   in the fault section. 
But if these effects are strong---as one might
guess would be the case for a weakly disordered
fault that is close to  planar---intermediate
size events will not occur and the behavior will
be qualitatively similar to the characteristic-
quake behavior observed in many faults \cite{EQ2}. Which behavior obtains would 
be determined by how the length scale above which events typically runaway 
compares with the length of the fault section.

Although this basic scenario of fault dynamics that was found in the simple 
heterogeneous fault model may have nothing to
do with what happens in the real earth, analog
laboratory systems might be found---or
synthesized---in which some of these ideas could
be tested. 

In these lecture notes, we have outined a framework for studying  the 
non-equilibrium "critical" behavior that occurs near the onset of macroscopic 
motion in many driven systems.  This enables us to understand  the origins, 
nature, and statistics  of avalanche-like events that can occur in these 
systems, as well as other qualitative and quantitative aspects of the critcal 
behavior.  But the examples given in this section have also illustrated a key 
point of these lectures: to be really useful, a phenomenological framework 
should be broad
enough and robust enough to show the roots of
its own failure. With judicious choice of which experimental systems to focus on 
-- and some luck -- this should enable enough
predictions to be made  that theoretical
results and underlying assumptions built
into models  are  falsifiable!

\subsection*{ACKNOWLEGEMENTS}

Whatever understanding the author might have of the problems discussed here is 
due in large part to interactions over the past fifteen years with students, 
postdocs and colleagues too numerous to mention.  To all of them, I am most 
grateful. I would also like to thank Jennifer Schwarz for comments on a 
preliminary version of the manuscript.
This work has been supported in part by the National Science Foundation via 
grants DMR 9106237, DMR 9630064, and Harvard University's MRSEC.


\begin{thebibliography}{99}
\bibitem{MFB} The velocity exponent, $\beta$, in mean
field theory is not fully universal. It depends on
whether or not the pinning forces are continuously
differentiable and, if not, on the nature of the
singularities in $f_p(u)$. For all but discontinuous
forces, this causes long time scales in mean-field theory associated with
the acceleration away from configurations that have
just become unstable. With short range interactions,
these should not play a role due to the jerky nature
of $\varphi(\vec r,t)$ caused by jumps. Thus the
mean-field models with $\beta_{MF}=1$ are the
``right'' ones about which to attempt an RG analysis. See \cite{Kar}
\bibitem{Aniso} If the system is not rotationally invariant, the depinning 
critical behavior can be very different; see the discussion in \cite{Kar}.
\bibitem{SOC} Whether critical behavior is considered "self-organized" or not is 
somewhat a matter of taste: if the systems we are considering are driven at very 
slow velocity, then they will be very close to critical. In another well known 
situation, when a fluid is stirred on large scales, turbulence exists on a wide 
range of length scales extending down to the scale at which viscous dissipation 
occurs. In both of these and in many other contexts the parameter which is 
"tuned" to get a wide range of scales is the ratio of some basic "microscopic" 
scale to the scale at which the system is driven.
\end{thebibliography}
\end{document}